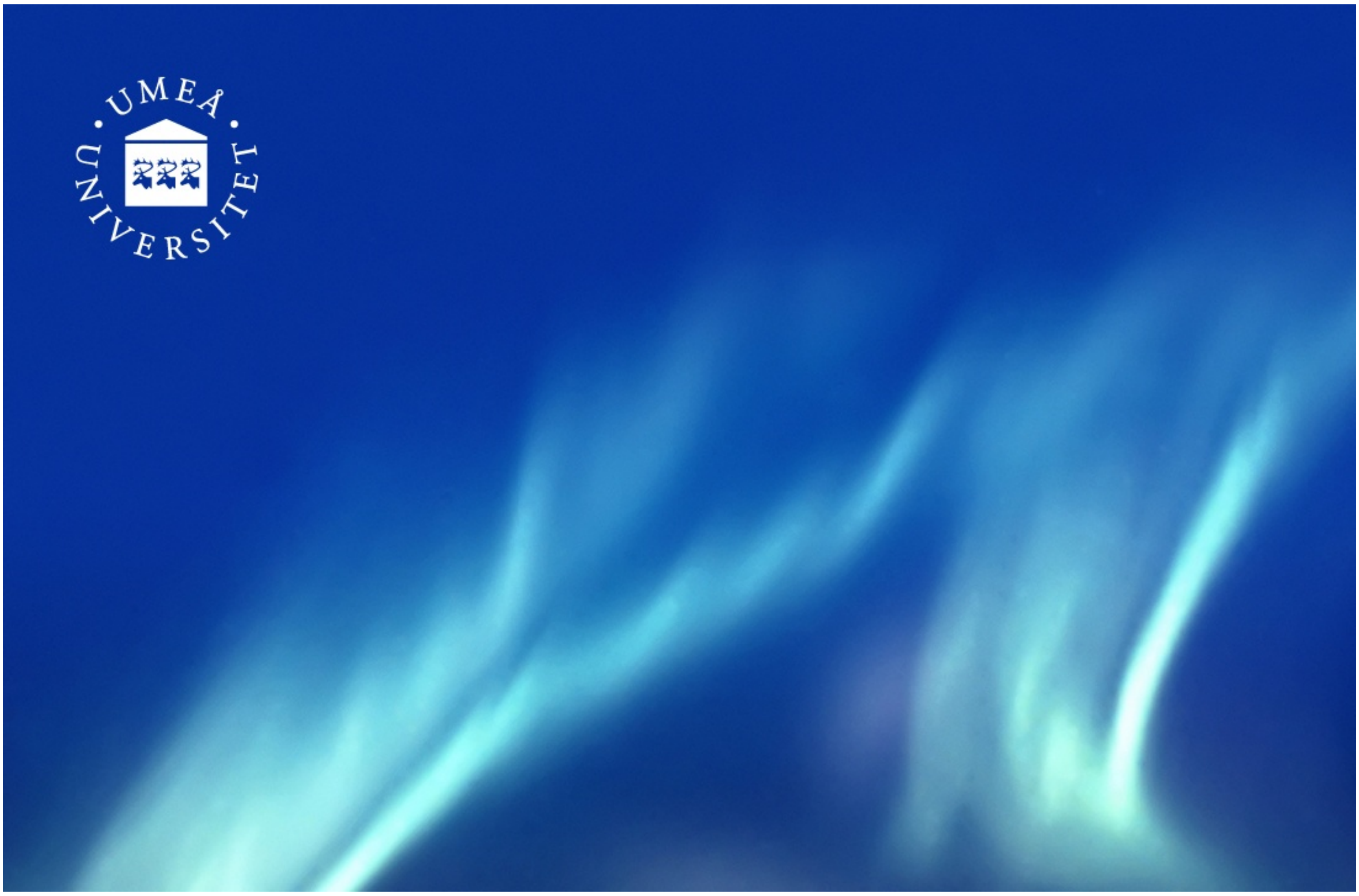

# Smart Spaces

## Challenges and Opportunities of BLE-Centered Mobile Systems for Public Environments

Tommy Nilsson

# Smart Spaces

## –Challenges and Opportunities of BLE-Centered Mobile Systems for Public Environments

## Abstract

*The application of mobile computing is currently altering patterns of our behavior to a greater degree than perhaps any other invention. In combination with the introduction of BLE (Bluetooth Low Energy) and similar technologies enabling context-awareness, designers are today finding themselves empowered to build experiences and facilitate interactions with our physical surroundings in ways not possible before. The aim of this thesis is to present a research project, currently underway at the University of Cambridge, which is dealing with implementation of a BLE system into a museum environment. By assessing the technology, describing the design decisions as well as presenting a qualitative evaluation, this paper seeks to provide insight into some of the challenges and possible solutions connected to the process of developing ubiquitous BLE computing systems for public spaces. The project outcome revealed the potential use of BLE to engage whole new groups of audiences as well as made me argue in favor of a more seamful approach to the design of these systems.*

# 1. Introduction

Probably no other technology in the history of mankind has caused such rapid and dramatic changes in our lives and abilities as the computer. During just the last few decades these devices have evolved from machines large as houses, used in a handful of military and research facilities, into relatively small and affordable tools present in nearly every household. Due to continuing erosion of cost and size related constraints of computing technology, mainstream consumers are today beginning to abandon traditional computers in favor of more versatile and lighter devices, such as smartphones. Consequently we are now in the process of leaving the desktop age behind us in order to embrace a more convenient paradigm in the form of mobile and wearable computing. Multiple studies are confirming this by showing that mobile platforms are starting to surpass traditional desktop computers in popularity (e.g. Büsching, Schildt & Wolf, 2012).

Today the implications of this development are however beginning to be felt far beyond just the proliferation of mobile devices. The reduced cost and size of computing technology has in turn allowed it to become increasingly ubiquitous. Computational capabilities are no longer exclusive to traditional or mobile computers, but rather being implemented to enhance the functionality of a growing spectrum of tools everywhere around us ranging from self-adjusting runner shoes all the way to fridges capable of downloading and displaying recipes (Kuniavsky, 2010). Moreover, by interconnecting these *smart things* with each other into larger systems, a whole new range of possible uses is being enabled. Essentially any artifact equipped with an identifier and wireless connectivity can today become a part of this emerging network, sometimes referred to as the Internet of Things or simply IoT (Golding, 2011). The concept of IoT has however its fair share of problems. Issues such as high power consumption and a resulting short longevity have traditionally been an important limiting factor holding back the process of equipping physical items with such wireless connectivity (Kamath & Lindh, 2012). As the latest generation of Bluetooth standard, aptly called Bluetooth Low Energy (henceforth BLE), is about to make entrance, hopes are high that these limitations might to some extent be addressed. Unlike its Bluetooth predecessor, BLE has a significantly reduced power consumption, with multiple sources claiming that a BLE transmitter can operate continuously for over two years using only a single coin battery (Kamath & Lindh, 2012). In combination with its small size, these factors could very well establish BLE as the dominant technology in the practice of granting physical items wireless connection and building a functioning IoT landscape.

Although in many ways replacing traditional computers, it would be a mistake to think of modern mobile devices simply as smaller and more convenient versions of their desktop counterparts. The development of mobile cameras and sensors, such as QR readers, gyroscopes or radio signal receivers, has granted the contemporary smartphones a substantial degree of context awareness. Whereas traditional desktop computers had to rely predominantly on data contained on their hard drives, our smartphones are increasingly reliant and responsive to information retrieved directly from our physical surroundings. In the emerging IoT landscape, one could thus argue that due to their mobility, context awareness and popularity, mobile devices such as smartphones constitute perhaps the most logical interface facilitating communication between people and these networks of smart objects. With close to 70% of the British population now owning a smartphone (Styles, 2013), it comes as no surprise that software companies are engaged in a fierce competition to produce new and increasingly powerful interaction interfaces drawing full advantage of these emerging technologies. Consequently, not only are software applications becoming increasingly deeply felt in an expanding spectrum of our daily-life activities, beyond just the traditional computer habitats, but we are now also witnessing the emergence of new types of applications opening up previously unavailable ways of interacting with physical spaces.

The possibilities of such technology have frequently served as inspiration for futurologist dreamers and forward-thinking scientists alike. Mark Weiser (1991), a pioneer ubiquitous computing researcher, has for instance described a vision of people and whole environments enhanced through computational resources that would provide information and services whenever and wherever desired. Although through innovations such as the BLE these dreams might now start to catch up with reality, we are still only beginning to understand the implications this constant immersion and interconnection through computation might have on our daily lives. Our actions performed in essentially any public environment, such as schools, hospitals, museums or shops could prove to benefit from the introduction of ubiquitous information technology (see e.g. Friedhelm & Raabe, 2011; González, Organero & Kloos, 2008).

These innovations and their ubiquity are inevitably pressing us towards rethinking the way we go about designing everyday experiences. Some designers are even arguing in favor of adopting new paradigms of interaction design based on widespread access to information and computational capabilities (Yvonne Rogers, 2004). There is a clear need to better understand technologies such as BLE in order to support and offer guidelines for future practitioners. One could argue that researching these emerging technologies is today not only relevant, but rather a necessity. Only through research will we be able to understand the opportunities they might present for future designers and ensure that their implementation into our everyday environments is done in a responsible and effective way that will result in spaces efficiently complementing and augmenting our human capabilities. Spaces that are, simply put, smart.

## 1.1 Problem Definition and Research Question

In a 2013 project the University of Cambridge Centre for Research in the Arts, Social Sciences and Humanities (CRASSH) explored the use of mobile technologies in museums in order to evaluate different approaches to enrich the way we engage with cultural information (Rosati, 2013). In 2014 both the University of Cambridge Museums consortium and the University's Computer Laboratory are attempting to build on this initiative by launching a joint project to examine the role of upcoming digital technology in engaging new audiences. As part of this effort, a mobile-based educational application centered around the use of BLE technology has been designed, produced and implemented into the university-owned Sedgwick Museum of Earth Sciences. The key ambition of this project could at the most basic level be described as finding an effective way of harnessing the capabilities of mobile and BLE technology to enhance the experience of museum visitors.

As a member of the research team, my task has been to assist with the design process as well as the formative evaluation. Although the project is being realized through the collaboration of roughly a dozen researchers and practitioners, its broad scope made it possible for me to pursue my own research goals within its boundaries. In line with the project's progress, I have attempted to explore the BLE technology to subsequently develop and critically examine a potential solution in the form of a mobile application facilitating the communication between users and BLE-based smart objects allocated in a public environment. By evaluating both the technical aspects and the usability side of this system, the ambition of my thesis was to answer the following research question:

- *In what ways could BLE centered mobile systems potentially improve our user experience in public environments and what are the approaches that might yield the best results when designing them?*

The broad scope set forth by this research question demands a firmly interdisciplinary approach to be taken. Nonetheless, the fact that the project deals with BLE has some important implications for the directions of the research to be presented. As has been argued, BLE is a brand new and still largely unexplored technology that is in fact not even available to the mainstream consumers yet. As such, very little material in terms of evaluation data and concept designs related to BLE is currently available. A significant portion of this paper will therefore have to focus on researching and exploring the actual technology by testing and evaluating its possibilities in the given context. At the most fundamental level, this thesis could thus be seen as a pilot study of the potential presented by Bluetooth Low Energy in public spaces, wherefore this paper will at least partially be drawing on the telecommunications research field. In order to answer the research question it is however not enough to simply find out if and how the technology is working. It is equally important that we ask ourselves whether its integration into public environments is a desirable goal from the end user point of view. To achieve this, besides including an extensive technological study, it is essential that a human-centered dimension is factored in as well in order to assess how to best improve users' experience in the given context through BLE. According to Herbert Simon (1996), the action of transforming current conditions (in our case the museum experience) into those that we prefer, is what lies at the heart of design practice. In line with this argument, it is feasible to say that another field related to the subject of this thesis is the design theory.

Finally, an additional factor that needs to be taken into consideration to accurately define the subject area of this research is the nature of public environments, and museums in particular. Few spaces are containing so many and so diverse media platforms as museum environments. Text labels, images, models and audio recordings are but a few of the information sources we can encounter in most museum spaces. This might seem unique, however truth be told, such diversity is present in public environments everywhere around us. Media artifacts such as advertisings, navigation marks or information sheets are all competing for our attention and contribute to the holistic experience mediated by a given public environment. In this sense, one could argue that visiting a public environment is by nature a cross media experience par excellence. A museum space could then in a way be seen as a microcosm of a *real world* public space. If we further factor in this diverse nature of public spaces, it is therefore feasible to further narrow down the subject matter of this paper as an area of cross media interaction design.

One issue that does make the museum context stand out from most other public environments is however the fact that most exhibitions contain some sort of educational element. In line with the CRASSH research initiative, our goal with the project was not to simply create a separate BLE centered experience that would only happen to take place in a museum. Rather the goal was to enhance the museum environment in a usable way by improving the existing experience. Since the experience offered by the museum hosting our project is essentially of an educational nature, learning and the possible role of computer technology in supporting education had to be factored in throughout this thesis as well. Likewise the particular nature of our museum had to be taken into account as well. In spite of this, the ambition of this paper is not simply to present an outline of an ad hoc solution. Although revolving around one particular museum environment, the primary concern of this paper is to highlight the problems and possible design solutions relevant to the process of integrating a BLE-based mobile system into public environments in general. As such the research conclusions brought forth aspire to be applicable to a broad spectrum of applications beyond just the museum setting. Subsequently, the process of finding an answer to the research question of this thesis is not only expected to present a coherent set of guidelines for future designers and developers dealing with mobile devices and BLE, but also provide insights contributing to the ongoing theoretical discourse on ubiquitous technology and its possible role in augmenting the human capabilities.

Although research into ubiquitous computing is increasingly common, the University of Cambridge's museums provide us with environments that are open to the public, without limiting our control of the implementation and evaluation. In combination with the university's resources and expertise, this opens up an unique possibility to perform a study that would not otherwise be possible. Moreover, as will be argued, studies of BLE are still scarce and focus predominantly on strictly technical aspects examined from the perspective of computer science. By exploring the use of systems centered around this technology in a public environment and by paying a particular attention to their design challenges and opportunities, it is my firm belief that this paper will fill an important gap in the previously described related fields.

Both the research question and the ideas presented in this thesis are my own and as such do not necessarily reflect the goals and opinions of the other team members. Likewise, unless otherwise stated, the design solutions and evaluations presented in this paper are the result of my own work.

The rest of this thesis will be organized as follows: The following section 2 will provide a brief summary of some of the research related to this thesis. Section 3 will describe the research project and its outcomes, including the technical evaluation, the prototyping and the user-centered evaluation. Section 4 will sum up the research findings through a synthesis with the previously described literature while attempting to offer an answer to the research question in the process. Future research directions will be suggested here as well. Finally, the paper concludes in section 5 by putting the findings into a broader context.

# 2. Related Research

As has been argued, the research question declared to be the objective of this thesis, as well as the context in which this objective will be pursued, require us to draw on research conducted in multiple areas. It is reasonable to say that finding and implementing a technological solution to a human problem is a broad and often not a very smooth process. Perhaps particularly so in those public environments that are traditionally not associated with the use of modern technology, as is the case in this project. Nonetheless it is generally argued (e.g. Davis, 1989) that the fate of every such process will ultimately be determined by the interplay of two basic factors: technological functionality and social acceptance. To begin with, the technology behind a solution must be functional, i.e. it must posses the attributes necessary to make it operational in the given context. Functionality alone is however no guarantee for success. A technological solution will not become successful merely because it is working, but rather because a sufficient number of target users is willing to accept it. Such acceptance will only be possible if the technology is adapted to comply with the relevant human needs. Whereas technical feasibility of a ubiquitous system, such as the one foreshadowed in the introductory section, is primary a concern of computer science and related engineering disciplines, social acceptance is somewhat more problematic to pinpoint to one specific academic area and is dealt with in design theory as well as various disciplines in the social sciences (e.g. cognitive psychology). To assess the relevant research it would therefore be wise to look into both of these broad areas.

As for the technological side of research activities, one area that is highly relevant to my research question and that has attracted a great level of attention, is the development and evaluation of solutions related to the IoT. For instance González, Organero & Kloos (2008) attempted to find the best way of utilizing ubiquitous computing to develop an infrastructure for innovative learning spaces. They concluded that mobile phones are versatile tools that in the near future will be able to support all the different types of processes needed in learning scenarios. This conclusion is perfectly in line with my aim to rely predominantly on mobile devices for facilitating communication between users and ubicomp devices. In relation to mobile platforms and their role in the internet of things, much focus has also been put on improving sensor technology required in the emerging internet of things media landscape. In pace with the technological progress, much of this research has naturally been shifting from traditional radio frequency identification (RFID) tags to near field communication (NFC) and most recently to the BLE (see e.g. Meydanoglu, E. S. B. & Klein, 2013; Kamath & Lindh, 2012).

In 2014 a group of scientists from the University of Cambridge, lead by Faragher and Harle (2014), conducted an extensive signal strength survey to better understand the capabilities presented by BLE in combination with mobile devices. As one of the very few existing studies of BLE, this survey is crucial for addressing my research question and will be referred back to throughout this thesis. At the same time, it is however important to note that whereas my paper attempts to evaluate BLE in relation to its potential use to enhance a human experience, this signal strength survey focused on assessing purely the technical aspect of BLE. Although the study confirmed the versatility of the BLE technology, a number of problems was also identified during the study. Most importantly, the measurements indicated a significant level of signal strength fluctuations. As a result, any attempt to use BLE for retrieving accurate spatial information, such as positioning, was argued to be futile.

Albeit somewhat discouraging, these findings were not unexpected. Ubiquitous computing is based on the principle of multiple distinct devices providing highly dispersed input, output and computational capabilities. Practically every such system implemented in public spaces, no matter how carefully designed, will always end up being a subject to variables beyond our control and as such cannot be expected to operate perfectly consistently. Due to the extreme diversity of environments where they are expected to operate, mobile devices are arguably suffering by these technical limitations to an even greater degree than other ubicomp systems. Patchy network coverage, fluctuating signal strength, deviations in positioning and the generally limited resources that mobile devices have at their disposal are but a small sample of the technical limitations users of mobile devices have to deal with. These deviations and discontinuities between what the system observes and what actually happens, are by some researchers described as *seams* (e.g. Weiser, 1994). Broll and Benfor (2005) argue that seams are traditionally seen as a detracting element of the user experience. As a result, designers are frequently putting significant effort into making the experience offered by a system as seamless as possible by masking over any inconsistencies. This is however often done at the cost of expensive investments into a better technology. In search of a more efficient solution, Broll and Benfor (2005) refer to the aforementioned pioneer of ubiquitous computing, Mark Weiser, who endorsed invisibility as the main design goal of Ubiquitous Computing. In Weiser’s view (1994), designers should hide the complexity and infrastructure of tools to prevent them from intruding on users' consciousness. Doing so will allow users to focus on their tasks and not on the tool itself. As an example of such invisible technology, Weiser (1994) names electricity, which we use, yet do not have to attend to. Paradoxically, according to him, such invisibility should not automatically be seen as an equivalent to seamlessness. He warns that making things seamless amounts to reducing all system components to their lowest common denominator and making everything the same. In his view, such approach would sacrifice the richness of every tool just to achieve a bland compatibility. Instead, Weiser (1994) suggests an alternative approach in the form of seamful systems *with beautiful seams* as the goal designers should strive for. In this sense, seamfully integrated parts of a system could indeed still provide a seamless interaction.

This view is echoed by Chalmers (2003) who argues that integrating seams into an experience is hard, but the quality of interaction can be improved if we allow each component to be itself. In line with these arguments Broll and Benford (2005) propose a new paradigm in the form of *seamful design*, revolving around the idea of revealing and exploiting technical limitations in ubiquitous computing. Since many of the seams arising when using a technology are in fact inevitable, instead of trying to hide them, they argue that we ought to use them to enhance the actual experience. In fact most of us are already using seams to enhance our experience, often without even knowing about it. For instance, we have all been in situations when we received a phone call we were not too eager to continue and dismissed the caller by claiming that we're driving through a tunnel and the signal is getting too bad to talk. In such simple situation, we are in fact using the common knowledge of an inconsistent signal coverage to our advantage.

In an attempt to demonstrate the viability of the seamful design approach, Chalmers (2003) developed a mobile location-based game called *Bill*, that was centered around the idea of exploring seams in wireless networking. Players were encouraged to collect virtual coins distributed on predetermined GPS positions. By finding areas with stronger signal coverage called “access points”, players could upload their coins to the game server in exchange for credits. On the other hand by staying out of these network covered areas, players avoided *pickpockets* threatening to steal their coins. Due to signal strength fluctuations, positions of the access points were dynamic and keeping track of them becomes an important part of the game. Chalmers concludes that by exploiting seams such as inconsistent network coverage and signal strength fluctuations as actual resources, aspects such as interaction, gameplay and usability were improved.

None of the researchers mentioned above does however claim that seamlessness is always bad, or that seamfulness is always good. Rather, by highlighting this underexposed and underused design approach, they are merely attempting to demonstrate that there is a lot of room for seamful work. When looking at existing related research, it seems undeniable that an overwhelming majority of relevant projects is indeed focusing on seamless design. This is perhaps particularly apparent in museum environments (e.g. Cameron & Kenderd, 2007). One prominent example may be the museum augmented reality guide developed as a joint project between the *Musée du Louvre* in France and the Japanese printing company Dai Nippon Printing (Arnaudov, Eble, Gapel, Gerl, Lieberknecht, Meier, Miyashite, Orclic, Scholz & Tachikawa, 2008) in an attempt to gain experience in innovative multimedia approaches that could be used to bring together visitors and artwork. By equipping selected artifacts with RFID tags and distributing ultra mobile PC's to its visitors, Louvre could provide every user with relevant commentary and animations triggered on the PC's simply by moving into the transmitting range of the RFID tag of respective artifact. Due to low complexity of the technology, seamless design proved to be working well. The perceived usability and overall experience was evaluated through observations as well as questionnaires and observations of the users. The study found that people paid greater attention to the artifacts and acted more in accordance with the route guide than if relying on more traditional technology, such as static PDA screens and audio commentaries.

Similarly positive observations were made by Jones and Jo (2004) who studied the use of ubiquitous technology in learning environments and argued that it enables students to access education more naturally. They went as far as concluding that ubiquitous computing may in fact be the new hope for the future of education.

Despite these findings, we have yet to see any widespread use of such technology in the museum context. It would be unwise to attribute this situation wholly to constraints such as cost of the systems. The willingness contra reluctance to embrace a technological solution is always influenced by a multitude of variables. For example Tang (2005) examined the use of Internet among 310 pre-service teachers and found that their attitude towards it was driven by factors such as support from their friends, their confidence level and not least by the perceived level of usefulness.

In 1989 Fred Davis pioneered the Technology Acceptance Model (TAM), originally developed to facilitate technology integration in business and help predict system usage. Davis argued that adoption of new technology depends predominantly on two external variables: Perceived Usefulness, defined as the degree to which a person believes that using a particular system would enhance her job performance, and perceived ease of use, defined as the degree to which a person believes that using a particular system would be free from effort. Demetriadis goes as far as pointing out that difficulties in using a technology can cancel out the benefits potential that users believe they will gain from its use (Demetriadis, 2003). In relation to learning spaces, such as museums, the TAM was applied and expanded by Timothy Theo (2008), who investigated teachers' attitude towards computer use. Theo claims that the perceived complexity goes hand in hand with perceived ease of use. He found that people were reluctant to use tools that appear to be difficult and conclusively argues that teachers have the power to transfer their believes and values to their students and their attitude towards technology is thus profoundly important.

While the TAM model offers doubtlessly interesting and relevant insights into the variables affecting our attitude towards technology, its practical application within the scope of the research question of this thesis is limited due to lack of any clear guidelines on how to shape the perceived ease of use and usefulness as well as other external variables such as the attitudes of people towards technology in general. Nonetheless it provides us with a basic framework through which we can better interpret the human-centered evaluation data presented in this thesis. The aforementioned seamful design paradigm is on the other hand providing us with potentially important clues regarding the design approaches that are efficient enough to be applicable when planing and developing BLE-centered public environments. As will be argued in the following sections, seamful design played an instrumental role in our concept.

# 3. Method

Before we proceed to the actual research project, a few logical considerations regarding its organization need to be explained. As has been stated in the introductory section, the objective of this thesis is to provide insights into the design and potential results of implementing BLE-centered mobile systems into public environments. Specifically, the location chosen for implementation and testing of our system was the Sedgwick Museum of Earth Sciences (henceforth Sedgwick museum). This mid size, one of eight university owned museums is located in central Cambridge and houses over 2 million fossils, minerals and rocks, covering nearly 4.5 billion years of earth's history. Its family friendly environment means that all major age demographics, including children, can be found among its avid visitors. Since one of the goals of our project was to find out how the common people in public environments would respond to BLE systems, it comes as no surprise that the broad spectrum of people represented among Sedgwick museum visitors made it into our testing ground of choice. Finding, analyzing and implementing a solution that would best comply with the needs of this environment was in essence a design project.

From a research point of view this might seemingly pose a problem. Design has traditionally been associated with solving of practical problems rather than making research contributions (Zimmerman, 2007). Using design as a research method is however not entirely without support from scholars. In fact there are claims that in some contexts design could even be seen as a preferable research method. For instance Rittel and Webber (1973) remarked upon the inability of system engineers to apply scientific methods to address social problems, such as urban crimes. Due to the conflicting perspectives of the stakeholders, there is simply no accurate way of modeling and addressing such cases using the reductionist approaches of science and engineering. They conclude that there is a need to solve these "wicked" problems through a design approach. In response to this need, Zimmerman (2007) argues in favor of adopting a methodology consisting of ideating, iterating and critiquing potential solutions, which would allow design researchers to continually reframe the research problem. In Zimmerman's view such *research through design* can generate several beneficial contributions. By undertaking problem reframing, designers can help identify gaps in behavioral theory and models. By evaluating the performance of prototypes and studying the impact they have on their environment, design researchers can also discover unanticipated effects as well as bridge a general theory to a specific problem space, context of use and a set of target users. In turn, this might contribute towards identification of opportunities for new technologies as well as advancements for existing solutions. Similar views were echoed by Fällman (2008) who proposed a model of research oriented design with a heavy emphasis on designing and building prototypes in order to demonstrate research contributions and in turn generate knowledge that would influence the design of commercial products.

In line with these arguments, the research brought forth in this paper relied predominantly on a design methodology as a tool to generate sufficient knowledge to answer the aforementioned research question. Specifically, the methodology consisted of three different stages (see figure 1). Firstly, as a consequence to the novelty of BLE technology and the general lack of previous research conducted in this area, my initial focus had to be on assessing its technological possibilities. Secondly, a concept design phase had to take place to translate my findings into a solution that would be best compatible with the human requirements in the given context. Finally, once I had a prototype of the solution successfully implemented, I had to conduct a human centered evaluation in accordance with the TAM to assess its perceived usability and general usefulness.

All these three stages of the design process had to be conducted in this given order. The remainder of this thesis will therefore be devoted to presenting each stage in turn, with particular emphasis on its role in the research methodology and the results it generated. At the same time it is crucial to keep in mind that all three of these stages deal merely with different aspects of the same problem, namely with finding a design solution applicable to the given context. The discussion section of this thesis will therefore be devoted to synthesizing all of the stages into one coherent answer to the research question.

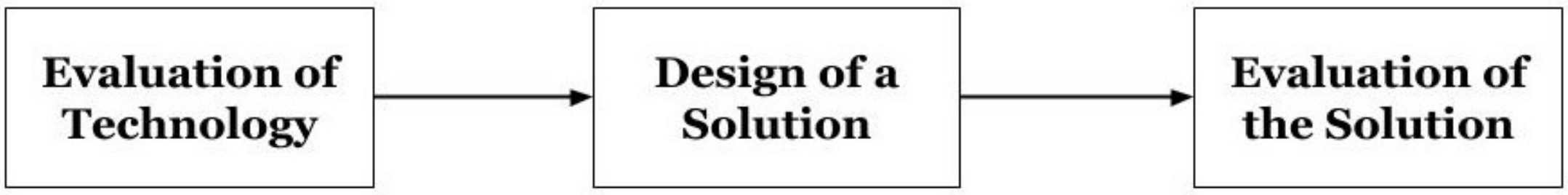

*Figure 1: Design methodology of the research project presented in this paper consisted of three broad stages.*

## 3.1 Technical Evaluation

BLE was designed as a low energy IoT platform and enables an invisible and unobtrusive way of interacting with physical objects in public environments, making it into a cornerstone of my project. One of my main concerns before the actual research was however that BLE's low power consumption would prove to have a negative impact on the transmission range when compared to traditional bluetooth. In order to methodologically assess the possible use of this technology, I therefore naturally had to start out with examining the distance across which BLE beacons were able to communicate with mobile devices. For this purpose an experimental study was deemed to be the most appropriate approach.

In order to conduct such experiment, I had to first develop a custom mobile application capable of measuring BLE signal strength. This ad hoc application was based on the simple principle of constantly scanning the surroundings for any BLE devices. Every time such device was detected and a connection with it was successfully established, the application displayed and kept updating a numerical value for the received signal strength (henceforth RSS) from that device. The application was built using the Android Eclipse development environment and written in Java.

During the actual experiment, a single BLE beacon was placed in turn at different locations in the Sedgwick museum along its main corridor (see appendix 1) at a height of approximately 1 meter above the ground. The beacon was set to emit signals at a rate of 10Hz. To account for any abnormal fluctuations in the signal received from the beacon, besides RSS, our mobile application had to additionally be programmed to measure standard deviation (henceforth SD). Each SD value was calculated based on 50 RSS samples (i.e. the total number of signal pulses received over a duration of 5 seconds). The smart phone Galaxy S3, running Android 4.3, was used as the platform for this application.

The actual experiment revolved around using the mobile application to measure changes in the RSS and SD values (i.e. the dependent variables) in relation to our location in the museum space (i.e. the independent variable). Furthermore, in order to account for the possible effect our facing direction could have on the received signal strength, each sample location in the museum space had the signal values measured while facing in turn north, west, south and east. These sample locations were distributed equally throughout the museum space. One dilemma I had to deal with was however the number of these sample locations, i.e. the resolution in which the museum space should be measured. It was clear that the higher spatial resolution would be used for the measurements (greater number of sample locations), the more accurate data would be produced. On the other hand, higher resolution would also require more time investment. In the light of multiple approaching deadlines, it was thus determined that measuring every square meter of the museum space would provide me with sufficient data to assess the BLE signal behavior.

Due to the relatively tedious and time consuming nature of this data collection process, five consecutive days had to be spent before the entire museum space was measured. The whole test was conducted at the Sedgwick museum during its opening hours. Hence, it is reasonable to assume that the performance of BLE in our test accurately reflected the dynamic conditions so typical for public environments (e.g. high layout complexity in combination with people consistently traversing the space). The whole experimental study was conducted with full knowledge and consent of the Sedgwick museum staff. Since the study didn't directly involve any human participants (beyond the research team), no potentially ethically problematic issues were encountered.

### 3.1.1 Results of Technical Evaluation

In order to make sense of the experiment results, the collected measurement data had to first be visualized. This was done by mapping each RSS value in the form of a color on top of a floor-plan of the museum. The colors signifying every value were ranging from green (high RSS) to red (low RSS). The resulting *RSS map* for each facing direction along with values collected from one of the BLE beacons can be found in appendix 1.

As expected, the study showed a steady RSS deterioration whenever moving away from a BLE beacon. Moreover, much like the aforementioned study by Faragher and Harle (2014), my experiment uncovered substantial inconsistencies in the RSS distribution across space, as well as irregular fluctuations over time. The nature of the museum interior with large shelves and artifacts allocated throughout the space proved to be an important factor contributing to the somewhat unpredictable signal coverage. Even the facing direction of the mobile user was frequently found to have a dramatic impact on the RSS. For instance, by turning my back towards the beacon, and thus effectively obstructing the path of the signal, the RSS could drop in some cases with as much as 10 dBm[1], which in this context is quite considerable. The museum visitors, seemingly randomly walking around in the space, were found to be another factor causing signal interference. In fact, at multiple occasions, larger crowds were able to block out the signal entirely.

These findings made it clear that any attempt of linking a set of RSS values to a specific position in the museum space would likely produce inaccurate results. Consequently, making a BLE-centered application operational on a context-aware basis would inevitably be problematic in this given environment. However, as has been discussed, these issues are an inseparable component of most public environments and as such must be factored in during the design process of any BLE-centered system intended for use in this context. As will be explained in the following section, unconventional approaches, such as seamful design, might hold the key to overcoming many of these obstacles.

## 3.2 Concept Design

As has been argued in the introductory section, mobile devices, such as smartphones, could potentially serve as ideal platforms for facilitating communication between users and a BLE enhanced public environment. This means that a mobile based solution using BLE technology to improve the user experience of museum visitors would then consist of two basic components: Firstly, a specific constellation of BLE beacons would have to be allocated in the museum space and secondly, a mobile application would have to be developed to

[1]dBm stands for Decibel-milliwatt and is an electrical power unit commonly used to define signal strength in wires and cables. (for more information, see Sjoquist, 2011)

communicate with this network of beacons in order to provide the user with desired information.

The technological evaluation has however manifested a number of trade-offs connected to the use of BLE technology. Most importantly, the range across which mobile devices and BLE transmitters can communicate was found to be highly dynamic and the use of BLE to facilitate any form of predictable experience is thus problematic. Conclusively, I argued that the fluctuating and inconsistent signal received from the beacons would most likely render useless any BLE-centered system providing users with information related to their position in the museum space.

It is feasible to say that these problems revealed by the technical evaluation are but a few isolated examples of a much greater issue, namely that the BLE technology cannot in the given context operate seamlessly. Due to existing technological limitations of BLE, there is simply no easy remedy to avoid the uncertainty caused by seams. Although there might be a way to design these seams out, this would likely be a time and resource consuming process. In our case one might thus argue that turning to unconventional design approaches, such as seamful design, and revealing the seams instead of hiding them, might be a preferable approach. Steve Benford (Chalmers, 2002) argues that there are four different broad approaches we might take to present seams: (i) a pessimistic approach consists of only showing information that is known to be correct, (ii) optimistic approach is centered around showing all information as if it would be correct, (iii) cautious approach explicitly presents uncertainty and lastly (iv) opportunistic approach seeks to exploit uncertainty. In my initial concept design I decided to go with the opportunistic approach, which in the words of Bill Gaver (1992) is "discordant, deliberately leading users to pause or reflect" and might thus be ideal for the museum setting.

The decision to adopt seamful design approach on its own would however not provide us with sufficient insight to leave the ideation phase of the design methodology and move towards prototyping of a concrete solution. Although the goals of the project upon which this thesis is based were primarily research oriented, we also had to follow certain directions and goals put forth by the University of Cambridge Museums consortium that were motivated by a desire of achieving customer satisfaction, rather than contributing to a specific research area. These requirements naturally had to be factored in during the design process and as such ought to be mentioned at least briefly.

First and foremost, the museums were interested in technological solutions that would attract new groups of audiences previously not interested in museum experiences. Young children were brought up as one such demographic group that is increasingly out of touch with traditional learning spaces and thus requires increased attention. In order to make our solution appealing to this type of audience, two criteria were deemed to be critical; the solution would have to be simple to understand and it would have to be entertaining. Achieving simplicity in the educational environment of the museum might seem slightly daunting due to the need of including at least some elementary learning value. I was nonetheless keen to design a solution that would enhance traditional museum visit without turning it into a science presentation. For this reason I attempted to avoid any unnecessary layer of complexity in the displayed information and relied predominantly on highly visual information such as illustrations rather than text to make the solution understandable to the younger audiences. Moreover, to make the concept entertaining, I decided to try to make it feel like a game.

Another requirement posed by the Museums consortium was that the system should preferably be applicable to all eight university museums. This meant that the concept had to be *expandable* in an organic way to tie together and establish a common user experience across all of the museum buildings. Furthermore the experience had to be designed to take into account that it could potentially be accessed through eight different entry points - each in the form of one of the university owned museums. At the same time I had to take into consideration that it is unlikely that most visitors will take the time to see more than one or two different museums. The solution thus also had to be fully functional and accessible by visiting only one of the museums.

### 3.2.1 The Prototype

With these requirements and limitations in mind I designed an experimental concept named *Lost Ghosts*. When walking through the museum space with the mobile application turned on, users would randomly encounter "ghosts" popping up on the screen of the mobile device and explaining to the player that they are lost and need help finding their way back to their home artifacts. Every such home artifact would be equipped with its own BLE beacon. While moving through the museum space, the ghost would then provide users with positive or negative feedback telling them if they're going in the right direction. The nature of this feedback depended on the RSS received from the BLE beacon of the home artifact. If the RSS value was strong, the ghost turned happy and provided the user with encouraging messages, such as "Yes, I can feel we're going into the right direction!". If on the other hand the user moved through an area with weak signal coverage, the ghost turned angry and reported that the direction is wrong (for visual illustration of the application, see appendix 2).

This means that the seam in the form of deteriorating signal strength over distance was used to estimate how far away from the artifact the user was. Fluctuations in the signal coverage contributed in turn towards making the search process more challenging and mysterious.

By carefully selecting the artifacts each ghost wanted to find and by making the ghosts contact users in a predetermined order, we could push the users towards visiting key artifacts in the exhibition in an appropriate order. This could in turn be used to mediate a narrative or any other experience that would require users to visit artifacts in a specific order.

Once all of these ghost-tasks are completed, the user receives an achievement for completing the museum and the option to share it through social media, such as Facebook. Finally, users are contacted by a last lost ghost, which this time explains that it comes from a different museum and needs help finding its way back home. This final ghost serves as an invitation to visit another museum space and makes users aware that every museum is a brick of a larger experience. The full concept is visualized in a flowchart seen in figure 2.

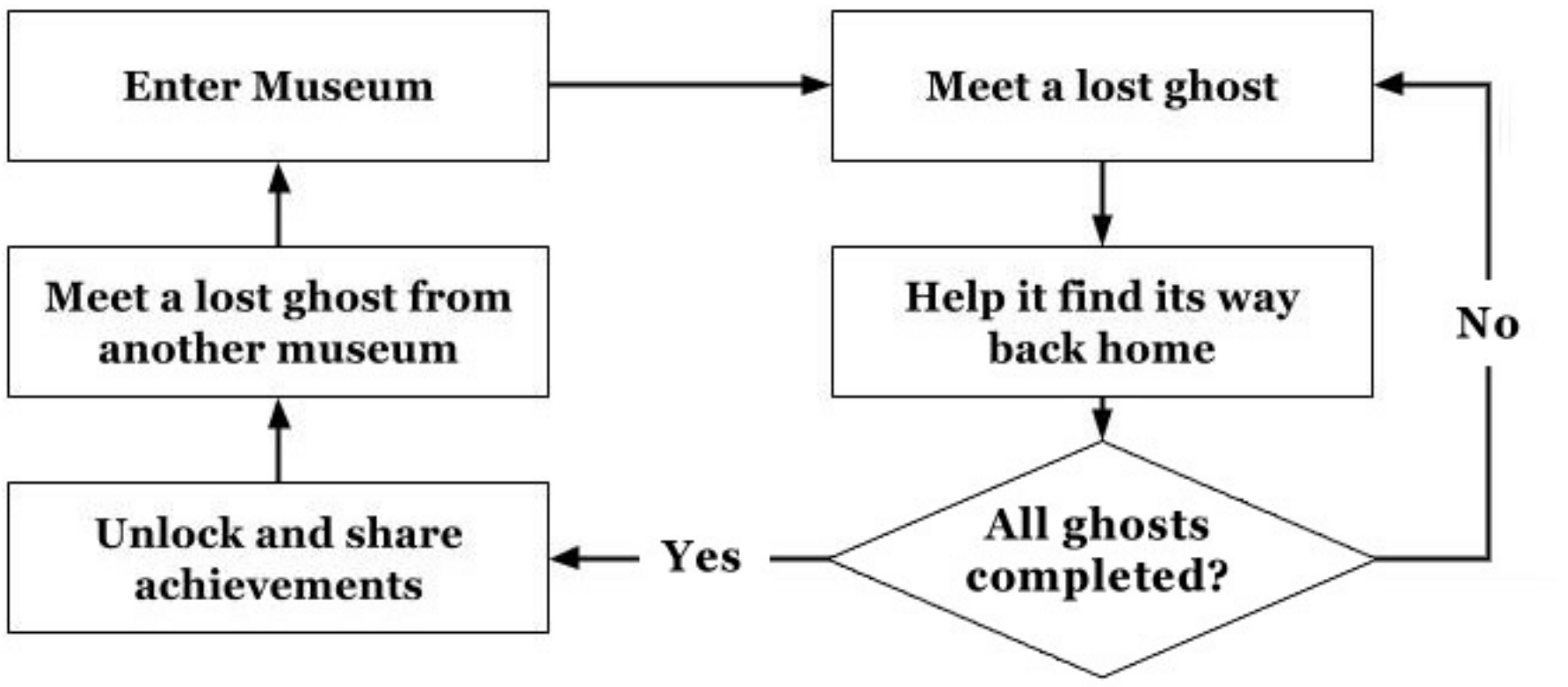

*Figure 2: outline of the Lost Ghosts concept*

This game-like application was deemed to be appealing to family visitors including children, who are normally perhaps not extremely excited about history or art appreciation.
A prototype of this application was developed using Unity 3D as the main development environment, primarily due to its ease of use and support of all major mobile platforms. The main scripting language was JavaScript. The system had to be tailored specifically to our given museum space, and the particular constellation of BLE beacons deployed in it. This means that the application is not functional outside of the Sedgwick museum space and for this reason will not be attached to this thesis.

## 3.3 Human-Centered Evaluation

Translating a set of design requirements and technological possibilities into a solution that may work in a given context is one thing. Answering broader questions related to the design approaches and the user experience implications of BLE-centered systems in public environments is however quite another. To even begin to address the research question stated in the opening chapter of this paper, I had to first turn towards a more human-centered evaluation involving real users.

The aforementioned TAM framework indicates that perceived usability and ease of use are two variables with a crucial importance when attempting to predict users' attitude towards a new technology. In accordance, my evaluation had to be based on examining how users perceived the system when interacting with it. By evaluating a prototype, my goal was primarily to assess how general public would react and perceive an experience which was presumably radically different from what most of them were used to from traditional museum visits. Doing so helped me in turn to assess the strengths, shortcomings as well as potentially risky aspects of both the seamful design approach as well as the use of BLE in general.

Due to the project being still very much underway and given the time constraints in place for the work presented in this paper, it was deemed unrealistic to conduct any form of summative evaluation through quantitative data collection. Under my circumstances, the large number of required participants as well as the time needed to collect and evaluate such data (Hartson & Pyla, 2012) would not be a feasible approach. Moreover, the data resulting from such evaluation would likely not produce an adequate scope of knowledge. Rather, in order to gain sufficient insights into the behavior, attitude, emotional response and overall perception of the system, a qualitative phenomenological approach (Creswell, 2013) was determined to be more appropriate. More specifically, the data was collected through an observation in the form of test monitoring in combination with a questionnaire distributed to the participants. The reason leading me to rely both on observation and on a questionnaire survey stemmed from the fact that although qualitative phenomenological approach is ideal for observing users' behavior while interacting with the system, mere observation would only show us what users are doing without answering the underlying motivations behind their actions. To give each participant the option to motivate their behavior and explain their impressions, the think-aloud method (Tullis & Albert, 2008) was utilized during the observation and questionnaires were distributed once the observation was completed. Relying on questionnaires alone would on the other hand run the risk of producing a certain level of bias in the responses, as participants could for example try to give the type of answers they believed were expected from them, without being entirely honest. Combining the methods described above was thus seen as the most productive approach.

Unfortunately, at the time of the evaluation, the previously described mobile application was still in a very early stage of development and as such was not fit to be used during the testing session. Instead, due to giving designers more flexibility in the early phases of development (Liu & Khooshabeh, 2003), a paper prototype was developed to serve as my main tool during the phenomenological evaluation.

This low fidelity prototype consisted of multiple screen views (see appendix 2), each printed on its own A4 paper and each displaying an important event in the game. The first screen introduced the user to a lost ghost which described the artifact in the museum it wanted to find. Test subjects were then asked to move through the museum space to find this given artifact. No further instructions were provided beyond what was stated on the paper prototype screen. Throughout the session I was walking behind the participants while observing their actions and based on their movement I supplied them with new screen views displaying negative or positive feedback. In order to simulate the fluctuating signal strength of BLE beacons, I used the previously developed signal strength maps of the museum space to determine what kind of feedback was adequate in various locations of the museum space.

Once completing the task, test subjects were asked to fill in a questionnaire assessing their impressions of using the prototype. The questionnaire consisted of 5 Likert scale questions, each in the form of a semantic differential spectra ranging between two opposite adjectives (i.e. *strongly agree* to *strongly disagree*). The main focus was on assessing the holistic user experience in accordance with the TAM theory by for instance asking users if they felt the need for additional help while using the prototype or if they would be willing to use a similar application more frequently. Additionally, 4 open-ended questions were included, giving the participants the opportunity to express themselves more specifically regarding their likes and dislikes about the system. These were later analyzed using the affinity diagramming method (Tullis & Albert, 2008) in an attempt to find any overreaching themes. The full questionnaire can be found in appendix 3.

The participants were conducting the evaluation one by one. During the observation part of the evaluation I collected observational data in the form of notes made on a traditional paper block. I was attempting to monitor the participants' actions, thought processes as well as other observations deemed relevant. Participants were encouraged to verbalize their thinking while performing the task and I paid particular attention to moments when their behavior expressed any emotional signs such as frustration or joy. Before receiving the questionnaires, participants were encouraged to be as open as possible in their answers and to not hesitate to give any form of critique if felt adequate.

The entire human centered data collection process took place during one weekend day, from approximately 10:00 to 13:00 in the Sedgwick museum with full knowledge and consent of the museum staff.

### 3.3.1 Illustrative Sample

As has been argued, the goal of the evaluation was to assess the usability issues as perceived by general users. An important aspect of the evaluation was thus selection of the test participants. One of the first decisions that had to be taken was regarding the number of users that ought to participate. Since the project did not yet enter a stage where it would be ready for a full summative evaluation, I decided a limited number would be sufficient. A common theory in usability testing, sometimes referred to as *the magic number five*, says that about 80% of usability problems will be observed by the first five test participants (Lewis, 1994). A similar view is echoed by Tullis and Albert (2008) who claim that using between 5 - 10 participants is sufficient to accurately assess the major usability issues. In line with these findings a total number of 6 participants was recruited for my evaluation. The test subjects were selected in a random manner from the visitors of the museum and roughly represented all major age demographics. The participants were selected with no regards to their past technological experience. Some of them could thus be seen as digital natives, whereas others had a degree of previous exposure to digital technology. I was not familiar with any of the participants before the test session, which meant that no personal relations could affect the answers given in the questionnaires, nor my interpretation of the observations.

Due to dealing with real people, unlike during the technical evaluation, possible ethical issues had to be taken into consideration this time. For this reason I saw it as vital not to withhold any information from the participants. At the beginning of the testing session, I paid a great deal of attention to introducing myself and the research project I was dealing with as well as explaining how the participants would contribute to my work. During this description, the word 'test' was deliberately left out when explaining the evaluation. This was to avoid putting any unnecessary pressure on the participants by making them feel like they have to perform well (Hartson & Pyla, 2012). They were all made aware that it is them who evaluates the system and not the other way around. Moreover, it was made clear to every participant that they are free to choose whether to participate or not and that they may quit at any time during the session. They were all notified that their answers would be kept anonymous. Additionally, where minors were present, their parents were asked for permission to let them take part in the research. Finally, before filling in the questionnaire, all participants were assured that their responses would not be taken out of context, and only used in the way previously specified.

### 3.3.2 Results of User-Centered Evaluation

The observation revealed that all of the participants understood the objective they were supposed to achieve and figured out how to achieve it merely by following the instructions provided by the prototype. With help of the positive/negative feedback, no participant failed to find the given museum artifact and thus successfully complete their task.

In spite of interacting with a paper prototype, participants had no problem approaching it as if it would be a real mobile device, while carrying it with them when walking around in the museum. This might be a good indication of the widespread social acceptance of mobile devices, with their use already becoming so deeply rooted in our culture that we approach mobile applications naturally without any hesitation. The only real difference compared to more traditional mobile applications, as perceived by the participants, was the context awareness of the app which demanded them to move around in real space and provided them with feedback relevant to their location in the museum. Nonetheless, they were all able to pick up the idea rather quickly. Multiple users even spontaneously begun suggesting innovations to the application, such as adding additional locations from the museum, that would be challenging to find and thus serve as excellent home locations of other lost ghosts.

Whenever doing a mistake, (i.e. moving away from the position of the hidden artifact) participants reacted appropriately to the negative feedback and changed their direction. Nonetheless there was one case where a participant experienced minor problems in completing the task. A child in the pre school age begun walking in the right direction and accordingly received positive feedback. In that moment he however turned around and begun walking in the opposite direction. This triggered negative feedback, which made him pause and reflect the situation for a brief moment. Eventually though he managed to figure out the correct direction and find the artifact. The fact that even a small child managed to complete the task, albeit after a brief struggle, was seen as testimony to the potential held by the system. Moreover, once locating the target artifact, the child manifested genuine joy at succeeding to complete the task. This degree of positive reaction was not observed on any of the adult participants. In fact many of them openly admitted that they find the experience be geared towards children.

The questionnaires largely confirmed my observations made during the test monitoring, with most of the answers being positive towards the application. An overwhelming majority of the participants responded that they did not feel like they needed any additional help while using the application. All of the participants also stated that they either *strongly agree* or *agree* with the statement that the application enriched their museum experience (full results of the questionnaire evaluation can be found in appendix 4). Furthermore, many participants stated that they found the application innovative, which was reinforced by the fact that all of the participants answered that they've never encountered any similar application before.

One potentially serious problem encountered during the evaluation was however that it quite often proved to be rather difficult to tell whether the opinions expressed by participants were referring to the actual BLE system or merely towards the mobile application (e.g. the graphical user interface, the ghost character, presentation of the objective etc. ). For example, one older lady praised the colorful graphical design of the application, which, albeit nice, was a largely irrelevant information with regards to the research question of this thesis.

In a way the greatest problem of the human-centered evaluation proved to be its novelty and the cultural climate in which everyone is perfectly used to traditional applications and treated my prototype accordingly. As such, it is recommended that the validity of the evaluation results presented in this chapter is treated with a degree of caution.

# 4. Discussion

The purpose of the research brought forth in this thesis was to examine whether the use of mobile based BLE systems might have a positive impact on our user experience in public spaces. This has in turn required us to first look into the capabilities of BLE technology as well as the possible design approaches that might prove effective when dealing with this technology in the given context.

As has been explained, the novelty of BLE meant that there was not much ground to built on in terms of previous research. One of the first extensive studies conducted on this topic was the aforementioned project by Faragher and Harle (2014) which provided a clear indication that the precision of BLE signal is currently not sufficient for advanced context-aware tasks, such as accurate positioning. The results provided by my technical evaluation did clearly support these findings.

Nonetheless, this per se does not automatically mean that BLE, at its present state, can not be used to build innovative experiences assisting us and enhancing our user experience in spaces of our daily lives. Albeit to my knowledge never applied to develop BLE-related applications, I attempted to argue that seamful design, as suggested by Broll and Benford (2005), might be an approach applicable to address some of the problems discovered during the technical evaluation. In his study, Oulasvirta (2004) argues that the process of seamful design consists of solving three main problems; (I.) Understanding of which seams are important, (II.) presenting the seams to the users and finally (III.) designing interactions with these seams. In our case the most obvious seams were constituted by signal strength falloff over distance in combination with unpredictable fluctuations arising as a consequence of spatial factors beyond our control, such as people walking around or the uneven and complex museum landscape. In the suggested design solution, I have therefore attempted to translate signal strength falloff into a distance estimate, with greater strength of the received signal being interpreted as the user being closer to the source BLE beacon. Although RSS fluctuations caused a partial distortion of this distance estimate, they also helped to make the distance feedback (i.e. comments by a *ghost* displayed on a mobile device) slightly more cryptic and in turn resulted in a challenging search game.

In terms of user experience, this meant that visitors were encouraged to actively move around and explore the museum space rather than remain static, the way we are used to from a majority of contemporary digital experiences. The participants in my user centered evaluation seemed to appreciate this by claiming that the application enriched their experience and that they would like to use similar applications more frequently. In fact the evaluation in general revealed that people are surprisingly open towards this new technology. Based on the TAM theory, I have attempted to argue that as long as potential users perceive the technology as being easy to use and as long as they are able to see some form of positive gain from using it, there is a substantial probability that the technology will reach a meaningful level of social acceptance. Based on the usability evaluation conducted in this thesis I do therefore believe that successfully communicating these two values to potential users ought to be seen as a priority for future designers working with ubiquitous BLE centered systems.

Although both the human centered evaluation and the testing of the BLE technology was conducted in the museum setting, I do not believe that these results should be looked upon as context-specific and non generalizable. Many challenges and attributes are shared across most public environments. The usability-related limitations of BLE technology encountered in the museum context are likely to be relevant in other settings as well. Likewise, due to the general nature of visitors, there is no reason to believe that the attitudes manifested by users towards the technology in the museum space would not correlate with those we would encounter in other public environments.

Nonetheless, the results presented in this study ought to be taken with a grain of salt. Particularly so because the scope and ambition of the usability test were rather limited. The fact that the system was still being under development and the limited sample of users taking part in its evaluation contributed to making the tasks conducted by the participants too trivial to pose any stress test for the concept and in turn cannot warrant any conclusive statement regarding the future mainstream potential of BLE-centered mobile systems in public environments. There is simply not enough data here to support such conclusion. Furthermore, I cannot fully dismiss the possibility of bias due to my long term exposure to the BLE technology in combination with the time and effort put into the design process of the system. It is arguably inevitable that this has colored my view at least to some extent.

Another fact that ought to be considered in relation to my research question is that this type of solution is per se nothing new. As mentioned in the second chapter, a similar system has already been successfully designed and implemented into a Louvre exhibition (Arnaudov et al., 2008). Unlike my own concept, the Louvre augmented reality guide was however a relatively seamless system centered around the use of simple RFID tags. I cannot therefore claim that seamful design is the only valid approach when developing systems for the given context. I do however feel inclined to maintain that the immense complexity and unpredictability of most public environments means that designers of all sufficiently advanced systems will have to factor in that no solution can be expected to operate perfectly consistently. Particularly so in the case of a technology as versatile and ambitious as the BLE.

Despite the different design approaches, both of these museum projects seem to point towards the conclusion that it is feasible to implement new technology into traditional public environments with a positive impact on the overall user experience, yet without damaging the original experience. For instance, by introducing museum visitors to a lost ghost trying to find its way to a certain artifact, they are essentially encouraged to perform the same set of actions they would normally do in a museum, except this time with the addition of various gamification elements. It is reasonable to believe that the base educational and cultural value of the museum visit thus remains intact. Moreover, my study suggested that by taking this approach, the user experience will be particularly improved when it comes to children. Since attracting new demographic groups was one of the main design goals of the project, these results might be interpreted as a success of BLE in combination with seamful design.

With regards to the research question of this thesis, it is therefore feasible to conclude that rather than providing a holistic assessment of the potential presented by BLE and the relevant design methods, this paper demonstrates the viability of a possible BLE-based interaction interface and the seamful design approach by carrying out an experiment with a prototype in a considerably constrained time and space. As such this thesis merely manages to scratch the surface of the BLE technology and the possibilities unlocked by its power.

## 4.1. Future Directions

The research presented in this paper is merely a pilot test of the possibilities offered by BLE in public environments designed with a seamful approach. Consequently, in many ways it rises more questions than it answers. Perhaps the most obvious future direction would be to develop a high fidelity prototype of a BLE system in order to collect deeper usability data and to improve the validity. (Liu & Khooshabeh, 2003) Designing and developing a similar system with adult users in mind could also provide us with a more accurate assessment of how a BLE experience might be received.

The BLE technology is still under development. It is reasonable to believe that future iterations of beacons, and other devices capable of transmitting BLE signal, might be more precise than those that were used for the evaluations presented in this paper. Alternatively, a more complex BLE system, with a more dense network of beacons, might help us to increase precision and produce more complex, yet precise, spatial data. For instance, constructing a system capable of accurate positioning through triangulation might be somewhat of a holy grail of BLE-centered context aware computing.

The fact that a BLE centered mobile system can be implemented into a traditional educational environment, such as a museum, definitely deserves a deeper reflection as well. Could the shift towards mobile and context aware computing have a broader impact on the way we go about learning? Could classrooms in some sense give way to educational experiences taking place out in the real world through interactions with physical objects using mobile platforms? If we at least briefly entertain the thought that classroom based education is not necessarily a paradigm set in stone, we might ask ourselves whether a highly mobile teaching experience, supported by modern technology, could in fact facilitate a more effective learning process. A comparative analysis of the learning results produced by BLE systems in comparison to traditional learning techniques might help us answer these questions.

# 5. Conclusion

The much quoted media philosopher Marshall McLuhan (1964) supposedly remarked that "we become what we behold. We shape our tools and then our tools shape us". The truth behind this message has perhaps never been more apparent than it is today. We are going through a unique time in mankind's technological history and just as the technology we are using changes, we are changing with it. The information technology explosion has greatly increased our interconnection and in turn our understanding of the world around us. Whereas the desktop age pushed us towards a state where our interactions with digital data occurred under relatively isolated conditions, such as while at work or at home, the shift towards mobile computing is making immersion with computing into a permanent state penetrating practically every aspect and environment of our daily lives. The real and the virtual, two entities that have existed relatively separately since the dawn of computing, are thus now becoming increasingly bridged.

More than two decades have now passed since Weiser (1993) made his prediction of a world in which people would be constantly immersed in digital information. Since then, technologies, such as NFC and various other RFID standards, have been successfully used to enhance our communication with real physical spaces. Due to its versatility, low power consumption and comparatively high data transfer rate, BLE represents arguably the most ambitious step towards this vision. Due to its support by all major smartphone platforms, BLE might well aspire to become the lingua franca of the newly emerging IoT media landscape and as such soon prove to lead the way for ubiquitous computing.

Regardless if enabled through BLE or other technology, in a world full of immersive experiences, being an interaction designer will no longer be about creating a good application, it will be about creating a good life. For rather than the ability of engineers to construct innovations, it is ultimately the actions of the designers that determines the direction and use of information technology. And perhaps in this lies the most significant morale surfacing as a result of the attempts described in this thesis. McLuhan had every right to stress the profound impact technology has on our lives. However we ought to never succumb to the feeling that we are at the mercy of technological progress. Although a technology may come with a certain set of predispositions, these are never arbitrary. Design is a mighty tool through which we can turn technological disadvantages into strengths, implement them into our systems, turn them into the driving elements of a good user experience and ultimately pave the way for a better use. And that is a powerful message indeed.

# 6. Acknowledgements

My sincere thanks go to professor Alan Blackwell from the University of Cambridge Computer Laboratory, for making it possible for me to spend my final semester in Cambridge, to experience one of the world's finest research communities and thus live out one of my wildest dreams. I'm very thankful to professor Andreas Lund from the Umeå University's Department of Informatics for the extraordinary patience he has manifested throughout the progress of this thesis while guiding me from across half of Europe (I will miss our skype conversations!). I wish to thank Dr. Carl Hogsden and Dr. David Scruton from the University of Cambridge Museums consortium for always supporting me and aiding me in my endless struggle with all the institutional bureaucracy. I'm grateful to Dan Pemberton, the collections manager from Sedgwick Museum of Earth Sciences, who allowed me to use his museum as a platform for the project and to whom I repaid by accidentally destroying the CCTV surveillance system of the entire museum building (sorry about that!). My sincere thanks also belong to Charith Perera from the Australian National University and to Jozef Mokry from the University of Cambridge for helping me out with the more technical side of application development. Further, I want to thank professor Rob Harle and professor Ramsey Faragher from the University of Cambridge Computer Laboratory for supplying me with the necessary hardware as well as tons of helpful tips during the technological evaluation and to Sarah-Jane Harknet for her invaluable help with the human centered evaluation. Finally, I'd like to thank my family for putting up with me throughout this mad adventure.

## Appendix 1 (technology evaluation)

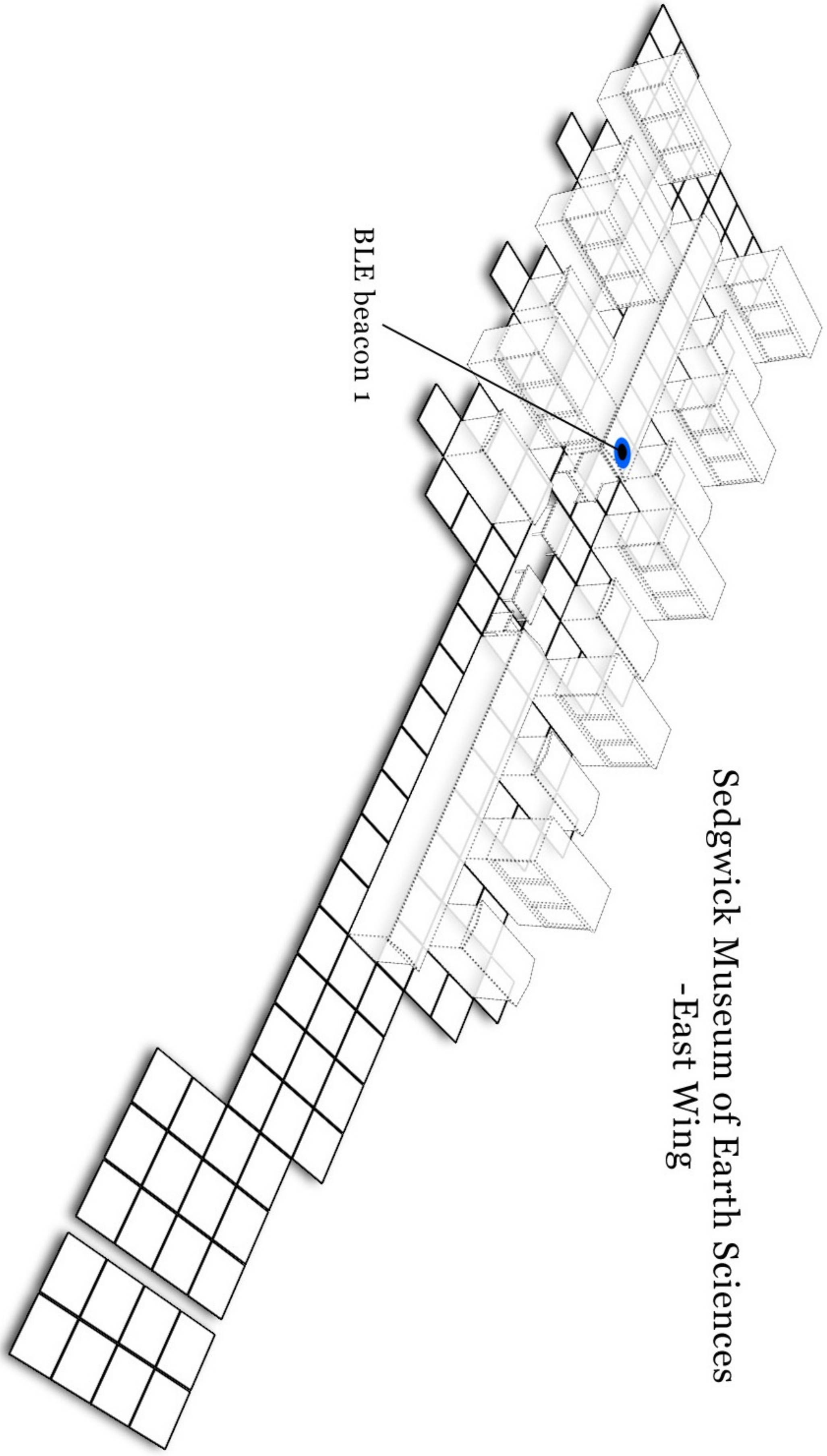

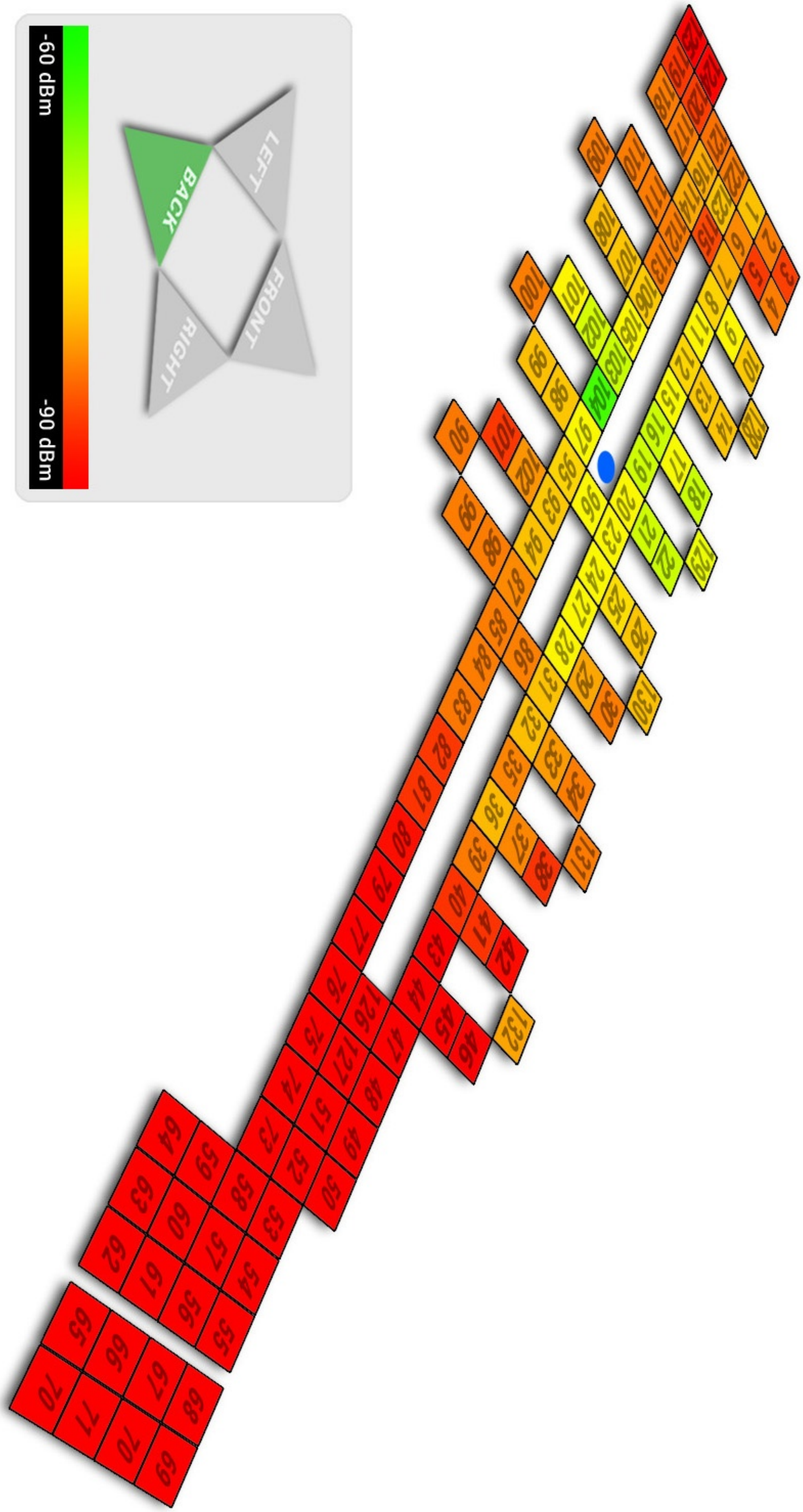

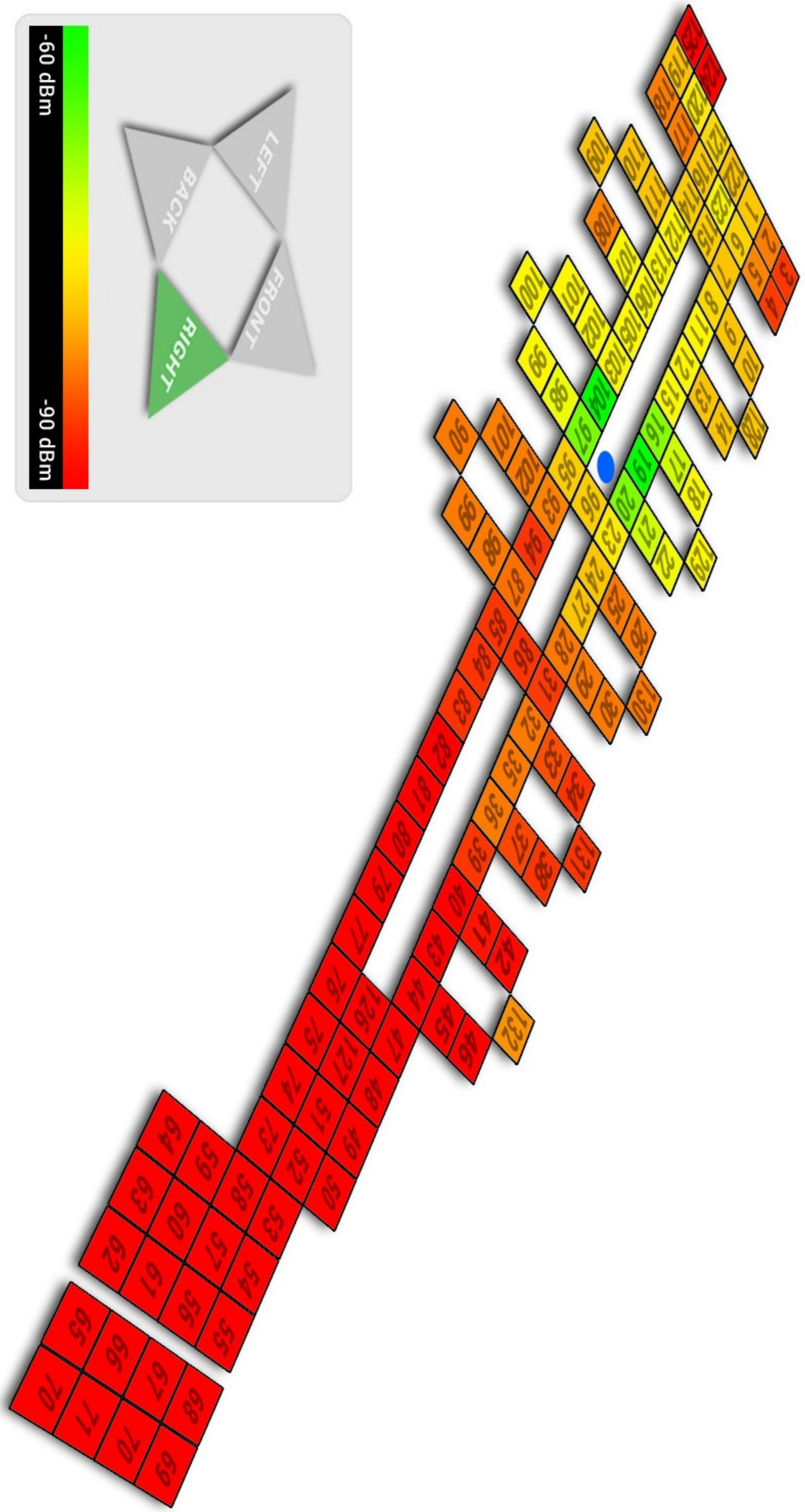

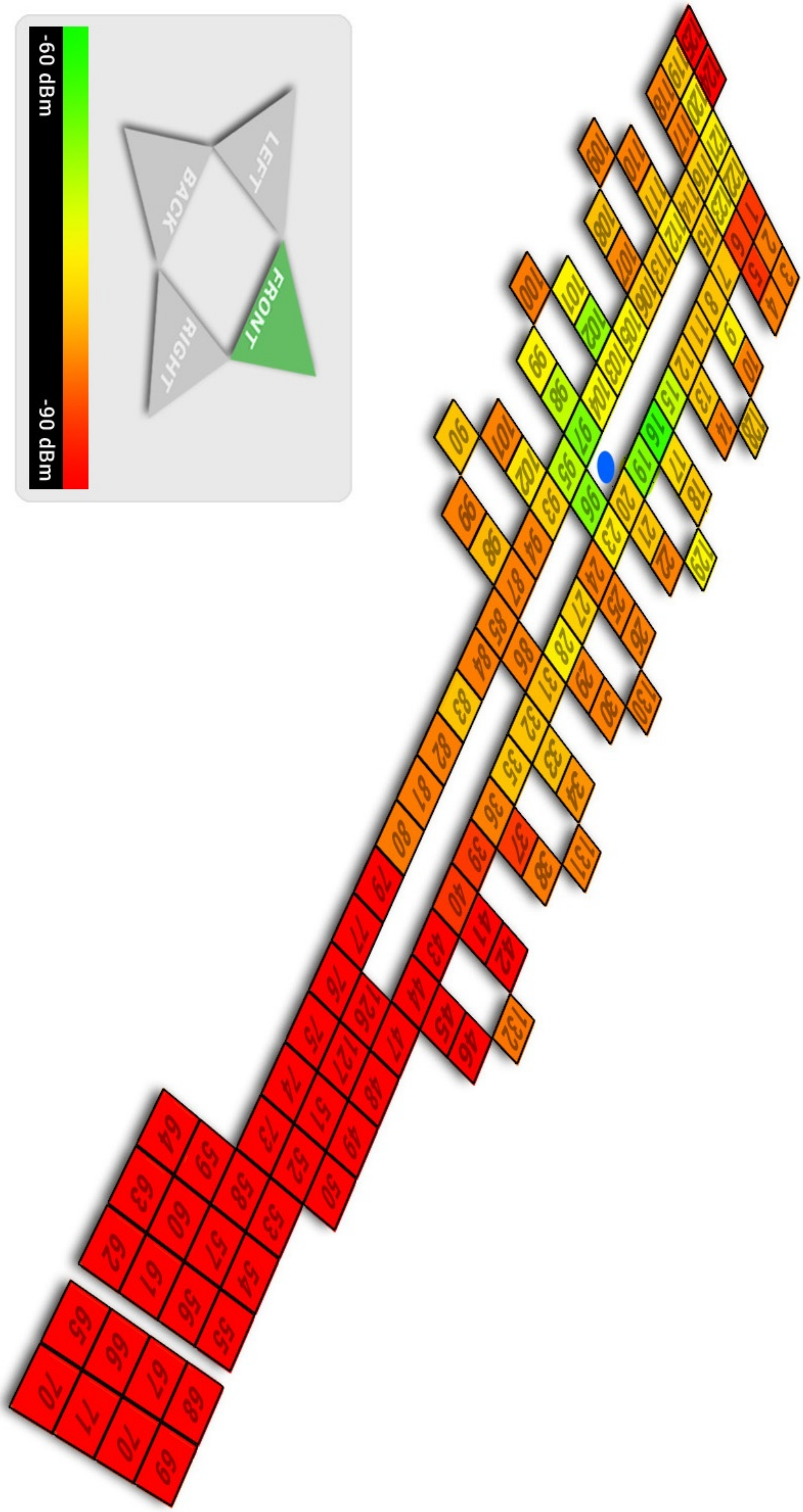

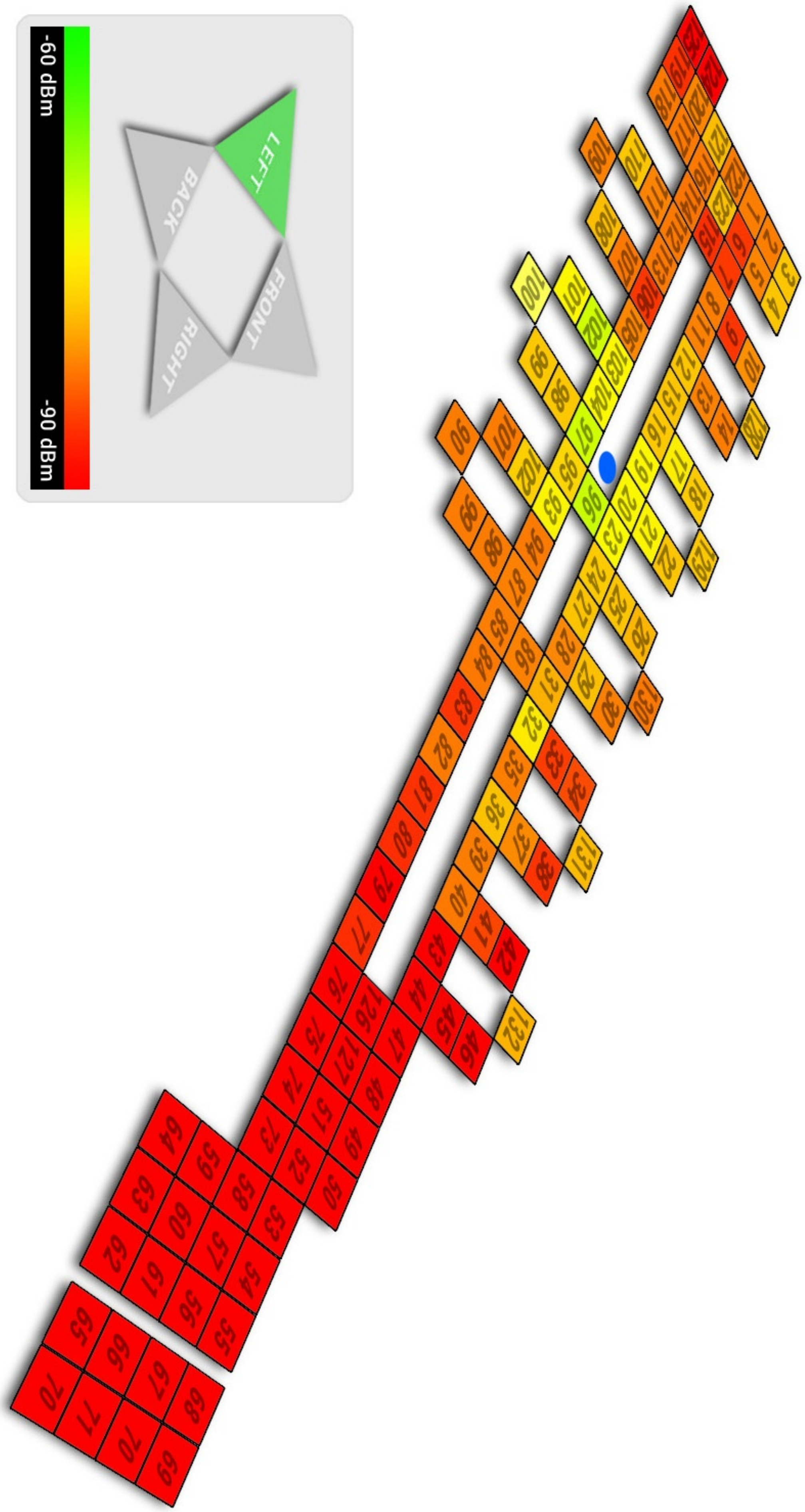

**East Wing - Beacon1**

| Location | BACK | | RIGHT | | FRONT | | LEFT | |
|---|---|---|---|---|---|---|---|---|
| | RSS | SD | RSS | SD | RSS | SD | RSS | SD |
| 1 | -81 | 5,5 | -80 | 1,7 | -88 | 2 | -85 | 2,8 |
| 2 | -84 | 3,6 | -84 | 2,7 | -86 | 2,7 | -86 | 4 |
| 3 | -89 | 2,2 | -88 | 3,2 | -83 | 1,7 | -80 | 2,8 |
| 4 | -86 | 3,7 | -88 | 1,4 | -85 | 1,4 | -82 | 1,4 |
| 5 | -88 | 3,3 | -84 | 2,2 | -89 | 1 | -87 | 2 |
| 6 | -85 | 3,2 | -79 | 1,4 | -81 | 1,4 | -85 | 3,6 |
| 7 | -79 | 5,5 | -82 | 3,2 | -82 | 2,7 | -88 | 2,2 |
| 8 | -80 | 5,1 | -82 | 3,2 | -80 | 2,7 | -85 | 3,3 |
| 9 | -78 | 4,3 | -79 | 5,1 | -77 | 2,5 | -87 | 2,7 |
| 10 | -82 | 3,5 | -80 | 3,9 | -84 | 3,9 | -85 | 3,5 |
| 11 | -75 | 2,7 | -75 | 3 | -81 | 5,4 | -85 | 4 |
| 12 | -82 | 3 | -75 | 4,2 | -79 | 3,5 | -81 | 4,1 |
| 13 | -82 | 1 | -82 | 2 | -80 | 6,2 | -86 | 3,3 |
| 14 | -82 | 4 | -80 | 2 | -84 | 3,6 | -83 | 2 |
| 15 | -78 | 5,4 | -77 | 3,3 | -73 | 4,4 | -81 | 5,3 |
| 16 | -74 | 5,8 | -69 | 3,2 | -66 | 2,2 | -81 | 5,9 |
| 17 | -76 | 4,4 | -71 | 3,6 | -80 | 3,9 | -75 | 4,4 |
| 18 | -72 | 2,5 | -76 | 3,3 | -80 | 3,5 | -80 | 4 |
| 19 | -71 | 4,6 | -63 | 1,4 | -68 | 5,7 | -75 | 3,4 |
| 20 | -76 | 3,9 | -70 | 1 | -80 | 5,6 | -77 | 4 |
| 21 | -72 | 3,3 | -73 | 5,1 | -82 | 3,3 | -77 | 4,4 |
| 22 | -73 | 2,8 | -76 | 1,4 | -84 | 4,4 | -79 | 2,7 |
| 23 | -77 | 2,8 | -75 | 2,5 | -79 | 3,2 | -76 | 2,8 |
| 24 | -75 | 3,3 | -81 | 2,7 | -83 | 3,3 | -79 | 2,8 |
| 25 | -80 | 5,1 | -85 | 2 | -85 | 4,8 | -79 | 3,5 |

| Location | BACK | | RIGHT | | FRONT | | LEFT | |
|---|---|---|---|---|---|---|---|---|
| | RSS | SD | RSS | SD | RSS | SD | RSS | SD |
| 26 | -79 | 2 | -85 | 3,7 | -86 | 3,3 | -80 | 2,8 |
| 27 | -76 | 3,7 | -82 | 6 | -81 | 3,2 | -81 | 2,8 |
| 28 | -77 | 4,2 | -84 | 2 | -78 | 2 | -83 | 3,9 |
| 29 | -84 | 3,3 | -84 | 3,6 | -84 | 3,6 | -81 | 2,8 |
| 30 | -83 | 2,8 | -83 | 2,8 | -84 | 3,2 | -85 | 2,5 |
| 31 | -80 | 4 | -90 | 1 | -79 | 3,6 | -79 | 3,2 |
| 32 | -82 | 4,6 | -86 | 5,2 | -81 | 4,5 | -78 | 2 |
| 33 | -85 | 3 | -88 | 2,8 | -89 | 2 | -91 | 1 |
| 34 | -86 | 2,7 | -89 | 2,2 | -87 | 2,7 | -89 | 2 |
| 35 | -85 | 2,7 | -86 | 6,7 | -81 | 3,2 | -84 | 3,7 |
| 36 | -80 | 3,6 | -86 | 2,1 | -86 | 2,7 | -80 | 2,8 |
| 37 | -88 | 2,7 | -89 | 1,4 | -87 | 2,7 | -86 | 3,6 |
| 38 | -86 | 3,3 | -87 | 3,5 | -84 | 1 | -88 | 2 |
| 39 | -88 | 3,7 | -89 | 2,7 | -87 | 2,2 | -84 | 2,7 |
| 40 | -84 | 3 | - | - | -88 | 2 | -84 | 2 |
| 41 | -89 | 3,5 | 90 | 1 | - | - | -89 | 2 |
| 42 | -82 | 2 | -82 | 2,5 | -81 | 4 | -82 | 2,5 |
| 43 | -78 | 4,1 | -77 | 4,4 | -78 | 4,6 | -79 | 4,5 |
| 44 | -81 | 3,2 | -86 | 3,3 | -85 | 3,2 | -83 | 3,7 |
| ... | | | | | | | | |
| 80 | -89 | 2,2 | - | - | -86 | 2,2 | -87 | 1,7 |
| 81 | -88 | 2,7 | - | - | -86 | 3,3 | -87 | 1,7 |
| 82 | -87 | 3,5 | - | - | -84 | 3,5 | -83 | 2,8 |
| 83 | -83 | 4,9 | -87 | 3 | -82 | 4,6 | -87 | 2,7 |
| 84 | -86 | 2,7 | -90 | 1,7 | -84 | 2,7 | -86 | 3 |

| Location | BACK | | RIGHT | | FRONT | | LEFT | |
|---|---|---|---|---|---|---|---|---|
| | RSS | SD | RSS | SD | RSS | SD | RSS | SD |
| 85 | -86 | 3 | -89 | 1,4 | -86 | 4,5 | -84 | 3,9 |
| 86 | -85 | 3,1 | -88 | 2,8 | -84 | 3,5 | -83 | 3,2 |
| 87 | -84 | 3,5 | -86 | 5,3 | -85 | 4,9 | -83 | 3,1 |
| 88 | -86 | 3 | -83 | 3 | -81 | 4,1 | -84 | 3,9 |
| 89 | -84 | 2 | -83 | 2,2 | -84 | 3,3 | -86 | 2,7 |
| 90 | -86 | 3 | -83 | 4 | -80 | 2,8 | -85 | 4,2 |
| 91 | -89 | 1,9 | -85 | 3,7 | -85 | 4,1 | -86 | 3,2 |
| 92 | -86 | 3 | -83 | 2,8 | -80 | 4 | -85 | 2,7 |
| 93 | -81 | 5,2 | -85 | 2,7 | -80 | 6,2 | -75 | 4,4 |
| 94 | -82 | 3,5 | -87 | 3,6 | -83 | 4,2 | -83 | 3,2 |
| 95 | -81 | 4,6 | -79 | 2,2 | -72 | 3,7 | -80 | 6,1 |
| 96 | -75 | 2,5 | -79 | 2,2 | -69 | 1 | -73 | 5,4 |
| 97 | -78 | 5,8 | -70 | 3,5 | -68 | 3,2 | -73 | 5,9 |
| 98 | -82 | 3,7 | -75 | 4,7 | -73 | 4,5 | -80 | 1,7 |
| 99 | -80 | 3,2 | -76 | 2,5 | -77 | 3,6 | -80 | 5,4 |
| 100 | -83 | 4,2 | -78 | 4,6 | -84 | 5,4 | -78 | 2,2 |
| 101 | -77 | 4 | -78 | 4,6 | -76 | 2,5 | -77 | 4,1 |
| 102 | -74 | 3,1 | -78 | 4,5 | -70 | 3,6 | -72 | 1,4 |
| 103 | -74 | 3,7 | -77 | 5,6 | -75 | 1,4 | -78 | 2,7 |
| 104 | -66 | 1,4 | -58 | 1 | -75 | 5,7 | -77 | 3,9 |
| 105 | -77 | 5,7 | -78 | 2 | -75 | 1,7 | -83 | 5,7 |
| 106 | -79 | 3,6 | -77 | 3,1 | -82 | 4 | -87 | 2,7 |
| 107 | -81 | 4,2 | -78 | 2 | -84 | 3,7 | -85 | 3,9 |
| 108 | -82 | 4,5 | -83 | 5 | -81 | 4 | -81 | 4,1 |
| 109 | -84 | 2,5 | -82 | 4,1 | -83 | 4,1 | -84 | 4,7 |

| Location | BACK | | RIGHT | | FRONT | | LEFT | |
|---|---|---|---|---|---|---|---|---|
| | RSS | SD | RSS | SD | RSS | SD | RSS | SD |
| 110 | -83 | 4,4 | -82 | 4,8 | -84 | 3,2 | -81 | 3,6 |
| 111 | -83 | 4,3 | -80 | 4,8 | -79 | 3,2 | -83 | 3,6 |
| 112 | -83 | 4 | -78 | 1 | -78 | 3,7 | -84 | 2,8 |
| 113 | -79 | 3 | -75 | 2,5 | -82 | 5,4 | -86 | 3 |
| 114 | -81 | 3,5 | -82 | 2,2 | -79 | 4 | -83 | 3,5 |
| 115 | -87 | 2,8 | -82 | 4,3 | -82 | 2,5 | -88 | 3,5 |
| 116 | -81 | 5,5 | -80 | 4 | -81 | 6,1 | -84 | 2,2 |
| 117 | -83 | 2,7 | -83 | 3,5 | -85 | 3,3 | -86 | 3,9 |
| 118 | -85 | 3,3 | -84 | 2 | -86 | 2,7 | -84 | 3 |
| 119 | -87 | 2,5 | -80 | 2,8 | -79 | 3,2 | -89 | 1 |
| 120 | -87 | 2,8 | -76 | 1,4 | -77 | 3 | -83 | 3,5 |
| 121 | -83 | 4,1 | -81 | 4,6 | -75 | 1,7 | -82 | 2,2 |
| 122 | -85 | 3,9 | -79 | 4,5 | -80 | 3,6 | -84 | 5,2 |
| 123 | -79 | 1,7 | -76 | 3,2 | -77 | 2,7 | -81 | 3 |
| 124 | -90 | 2 | -91 | 2,5 | -90 | 4 | -90 | 2,5 |
| 125 | -90 | 4,1 | -91 | 4,4 | -91 | 4,6 | -91 | 4,5 |
| 126 | -90 | 3,2 | -89 | 3,3 | -90 | 3,2 | -89 | 3,7 |
| 127 | -91 | 2,5 | -90 | 2,7 | -91 | 3,1 | -90 | 2,5 |
| 128 | -81 | 4,9 | -79 | 3,2 | -84 | 3,3 | -82 | 4,9 |
| 129 | -76 | 3,2 | -80 | 2,2 | -75 | 2,7 | -79 | 2,9 |
| 130 | -79 | 3 | -83 | 2,7 | -84 | 3,1 | -86 | 3,2 |
| 131 | -87 | 4,9 | -89 | 4,2 | -88 | 4,2 | -81 | 2,5 |
| 132 | -85 | 5,2 | -85 | 4,1 | -88 | 3,9 | -81 | 4,8 |

# Appendix 2 (Lost Ghosts prototype)

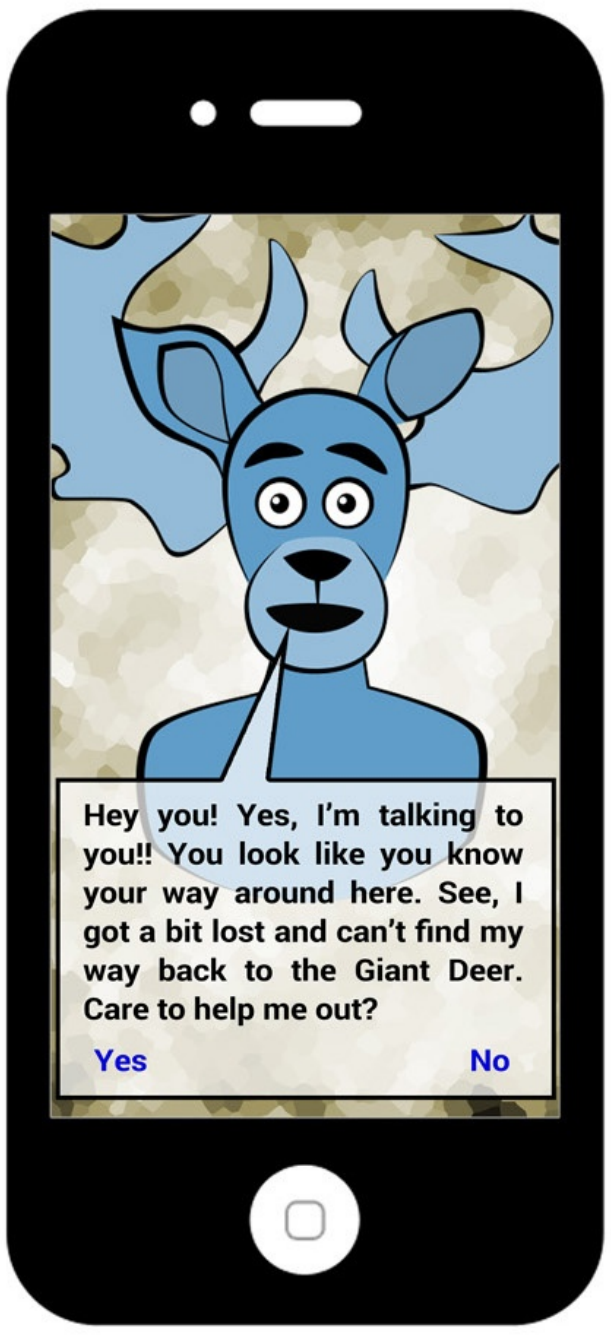

start

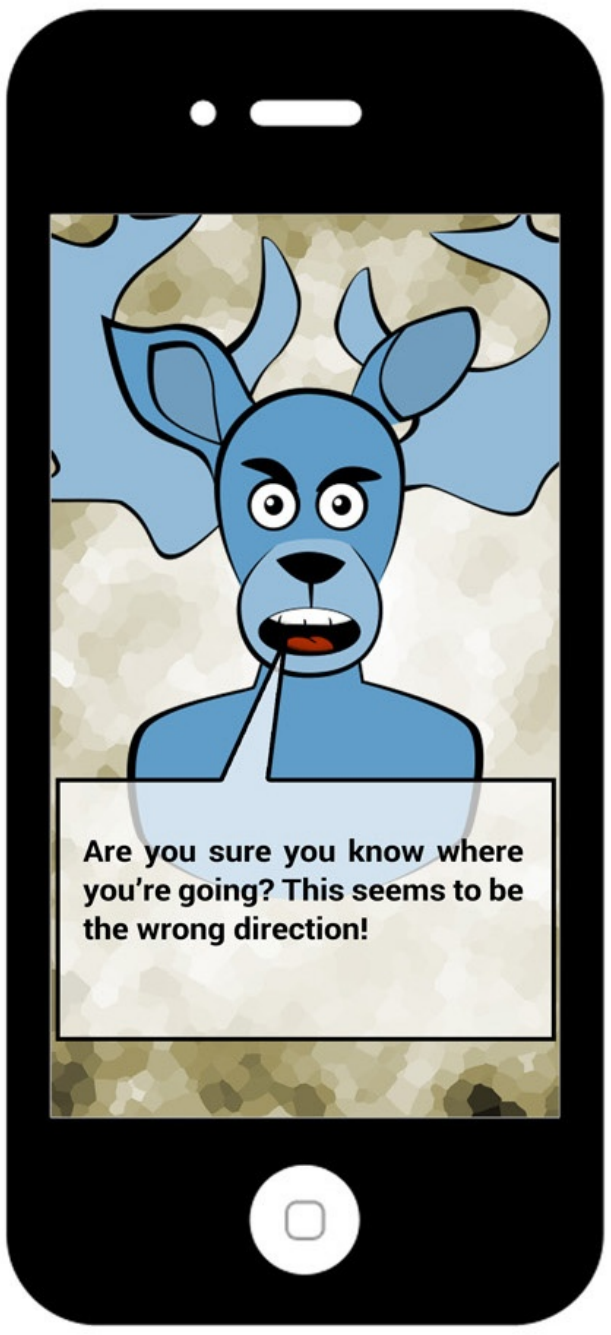

negative feedback

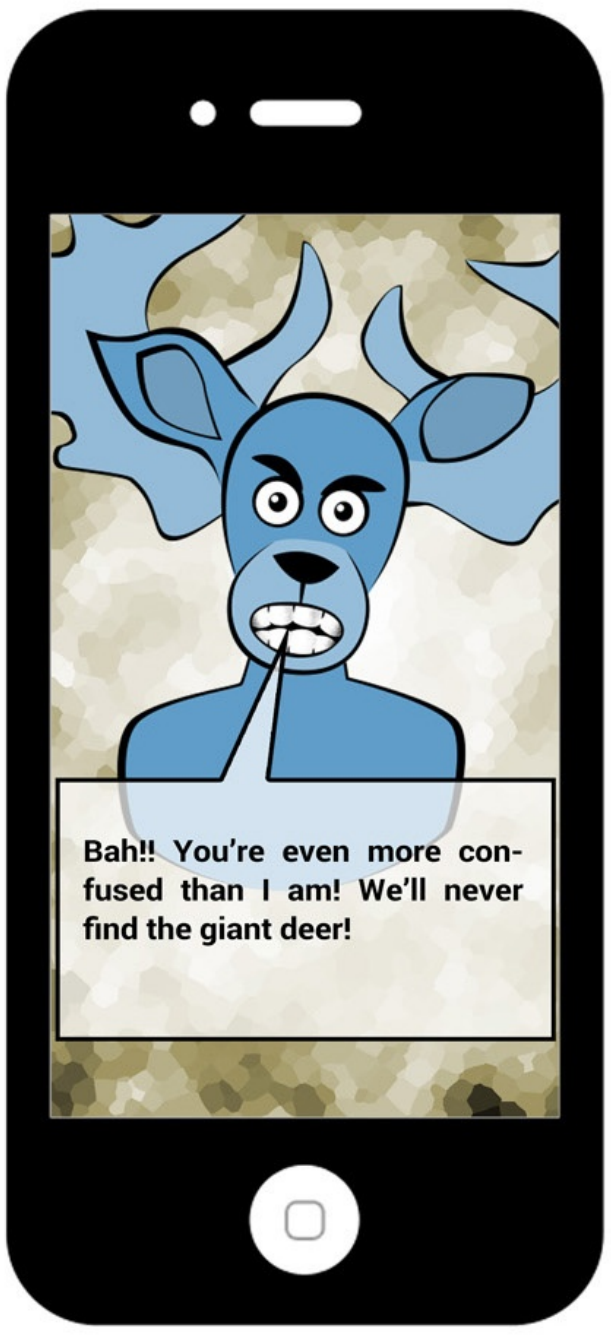

very negative feedback

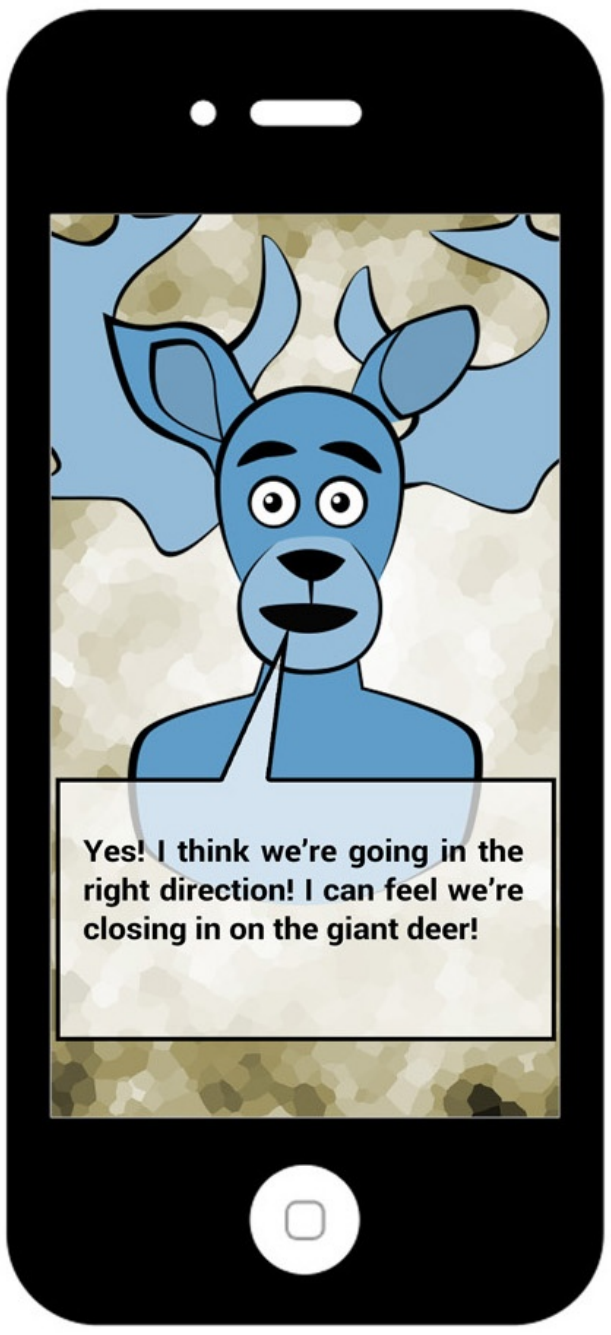

positive feedback

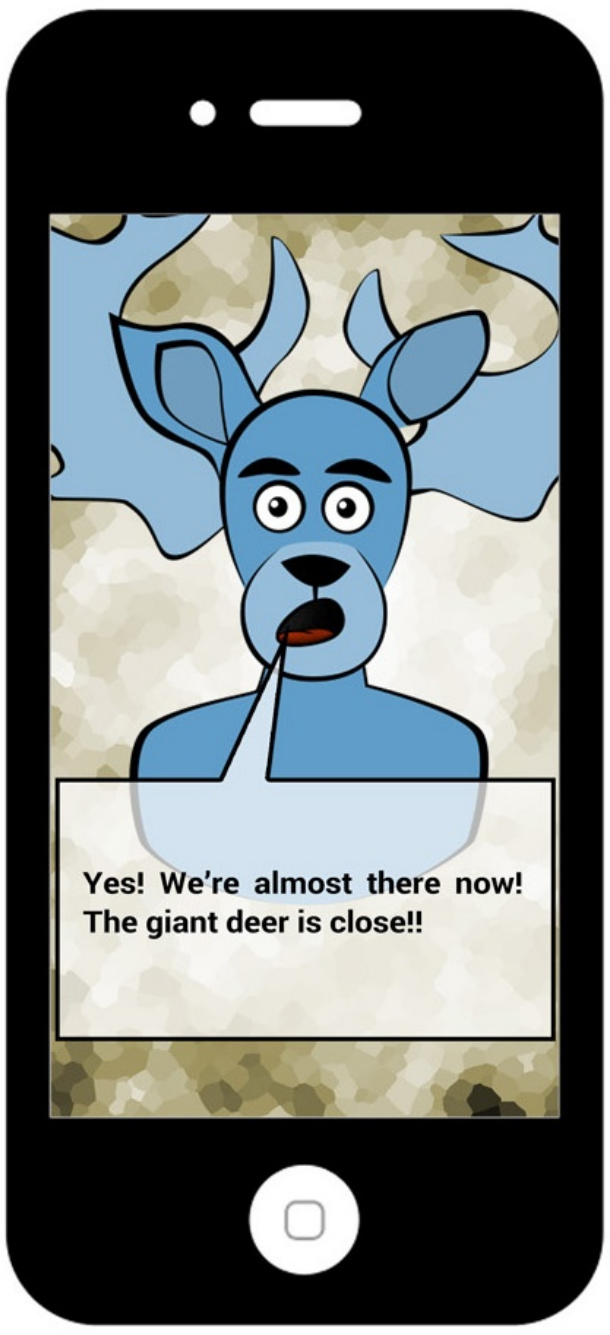

very positive feedback

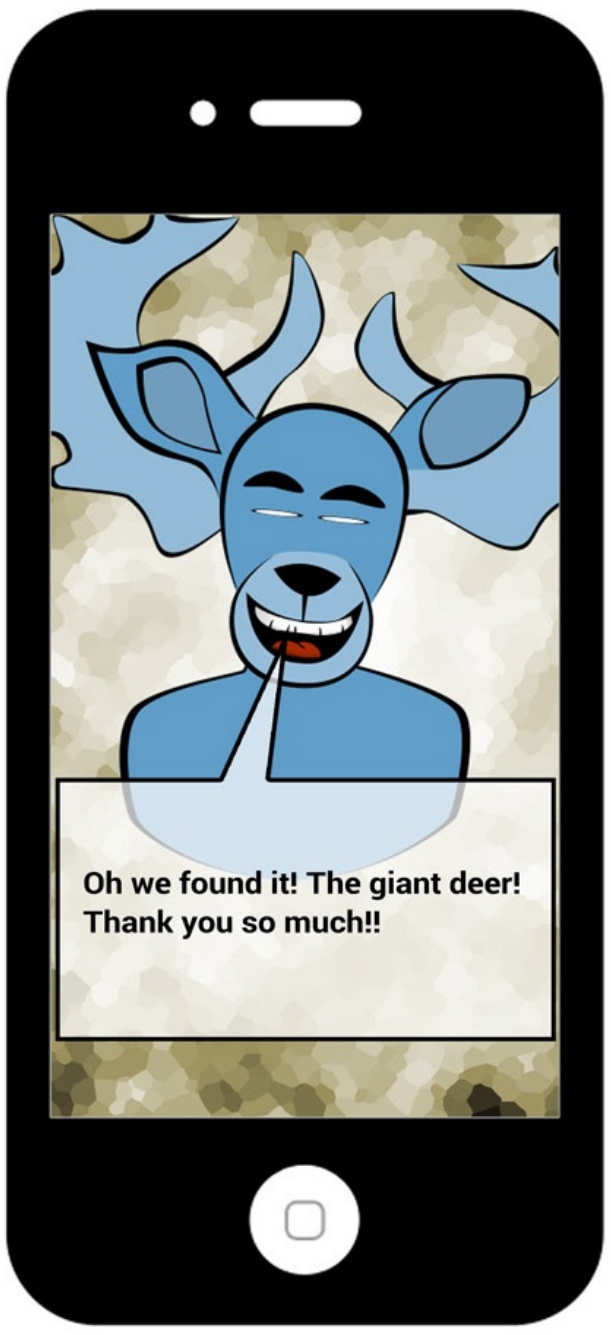

end

## Appendix 3 (questionnaire)

# Questionnaire - Museum-based Bluetooth LE Applications

Sedgwick Museum of Earth Sciences, May 17, 2014

**To what extent do these statements correspond to your own opinions? Please circle the best option.**

**1 - I found the application easy to use:**

strongly agree    agree    neutral    disagree    strongly disagree

**2 - I think most people would learn to use this application rather quickly:**

strongly agree    agree    neutral    disagree    strongly disagree

**3 - I think the application enriched my museum experience:**

strongly agree    agree    neutral    disagree    strongly disagree

**4 - I would like to use this kind of application more frequently:**

strongly agree    agree    neutral    disagree    strongly disagree

**5 - I never felt the need to ask for help:**

strongly agree    agree    neutral    disagree    strongly disagree

**6 - What did you like the most about the application?**

**7 - What did you dislike the most about the application?**

**8 - Is there something you would like to add/change in the application?**

**9 - What similar games/applications do you know?**

# Appendix 4 (Questionnaire answers)

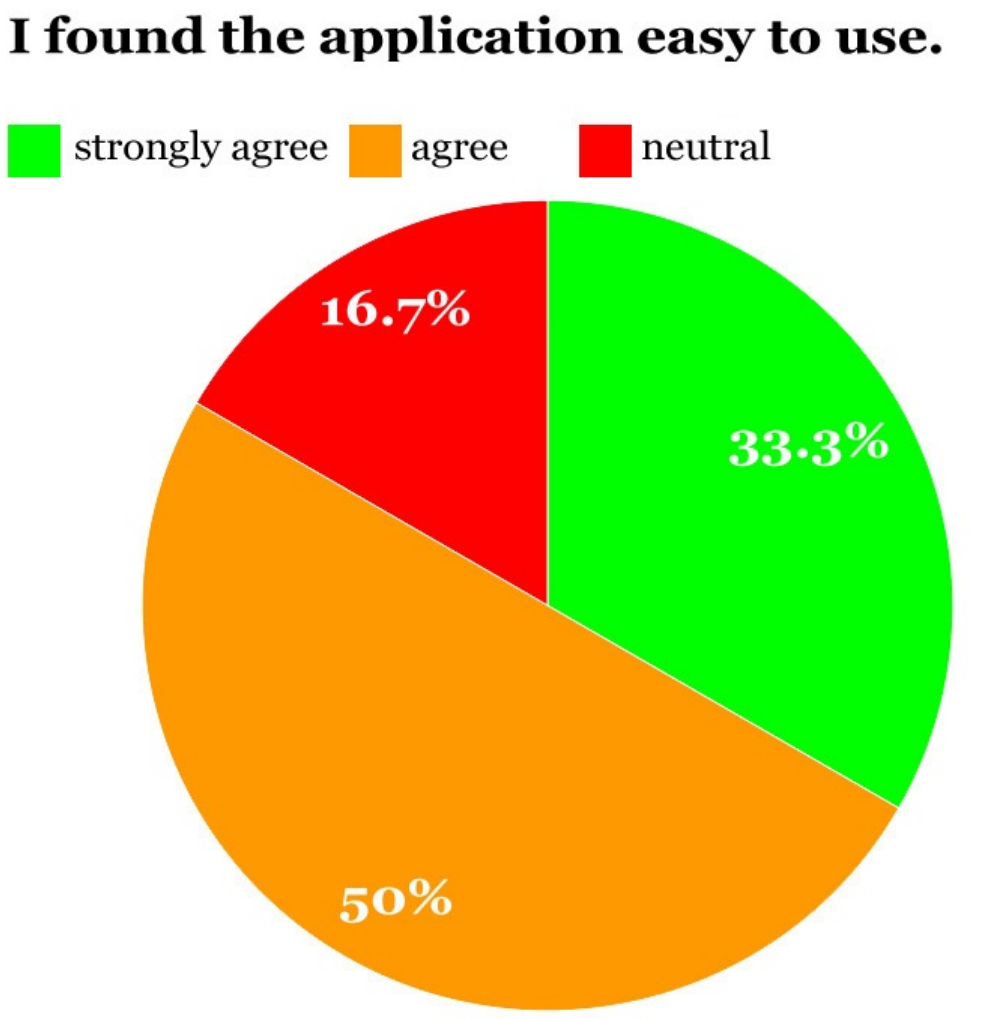

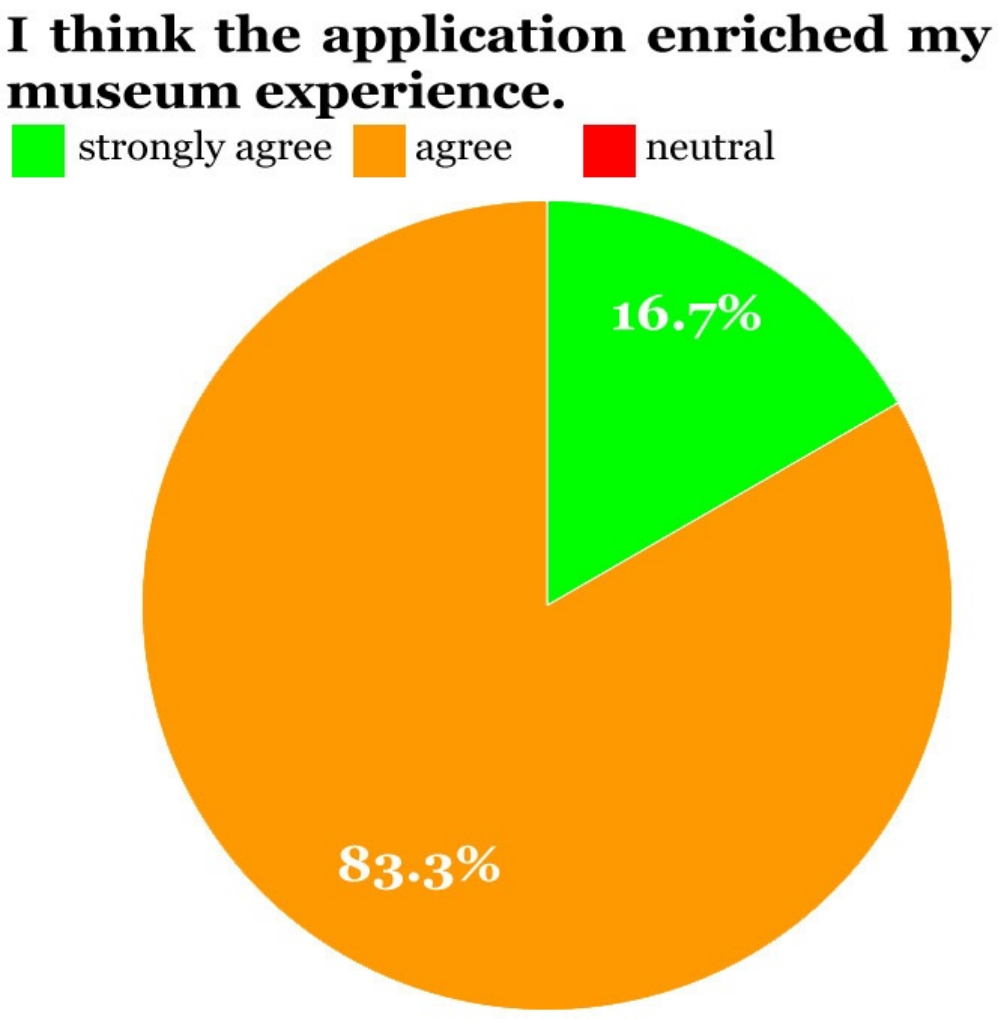

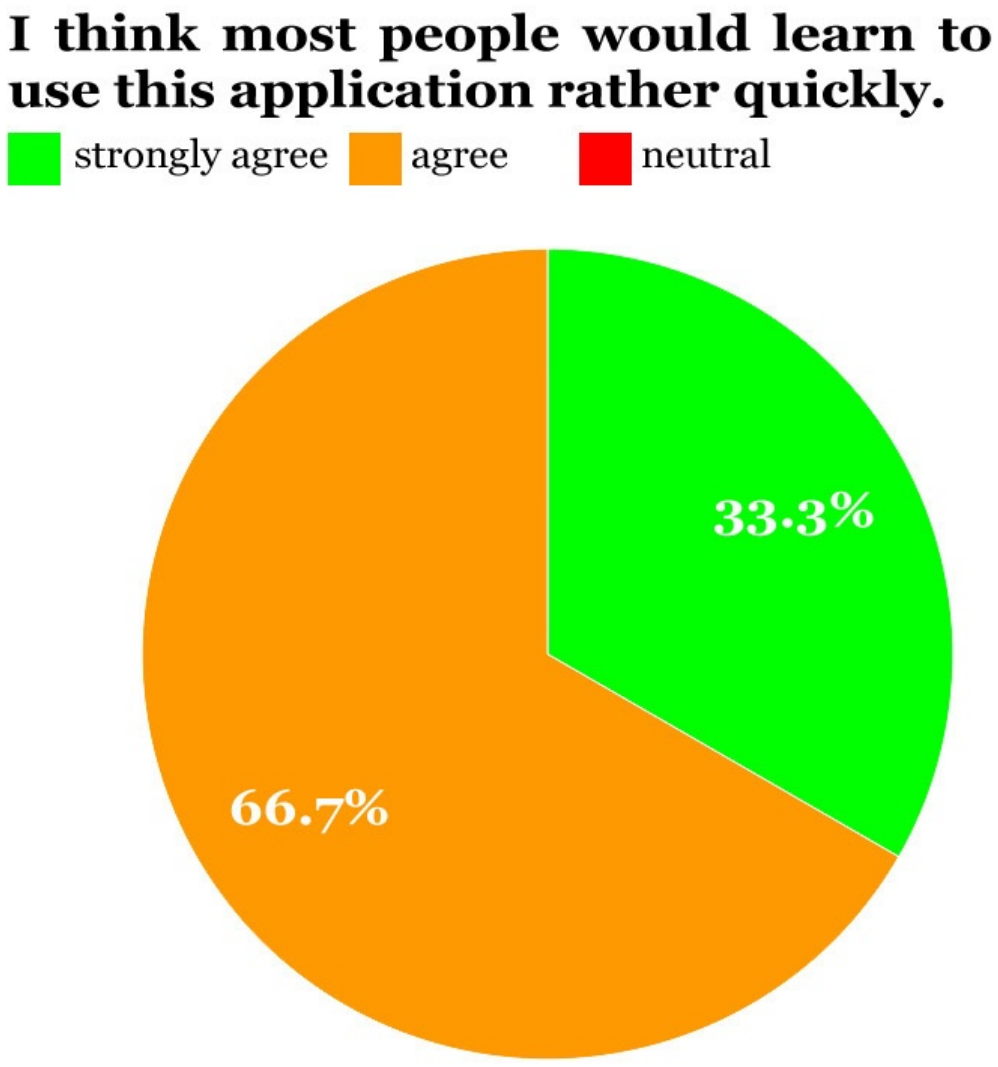

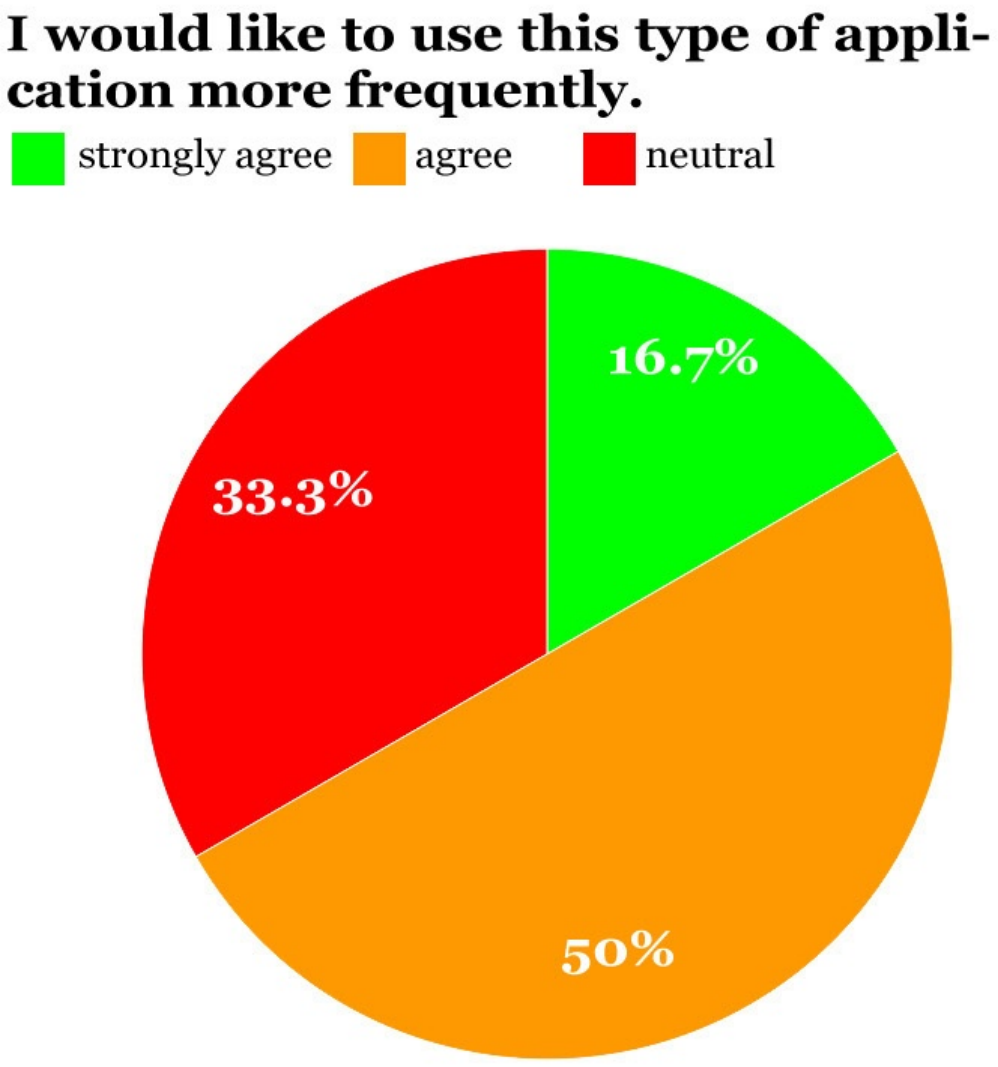

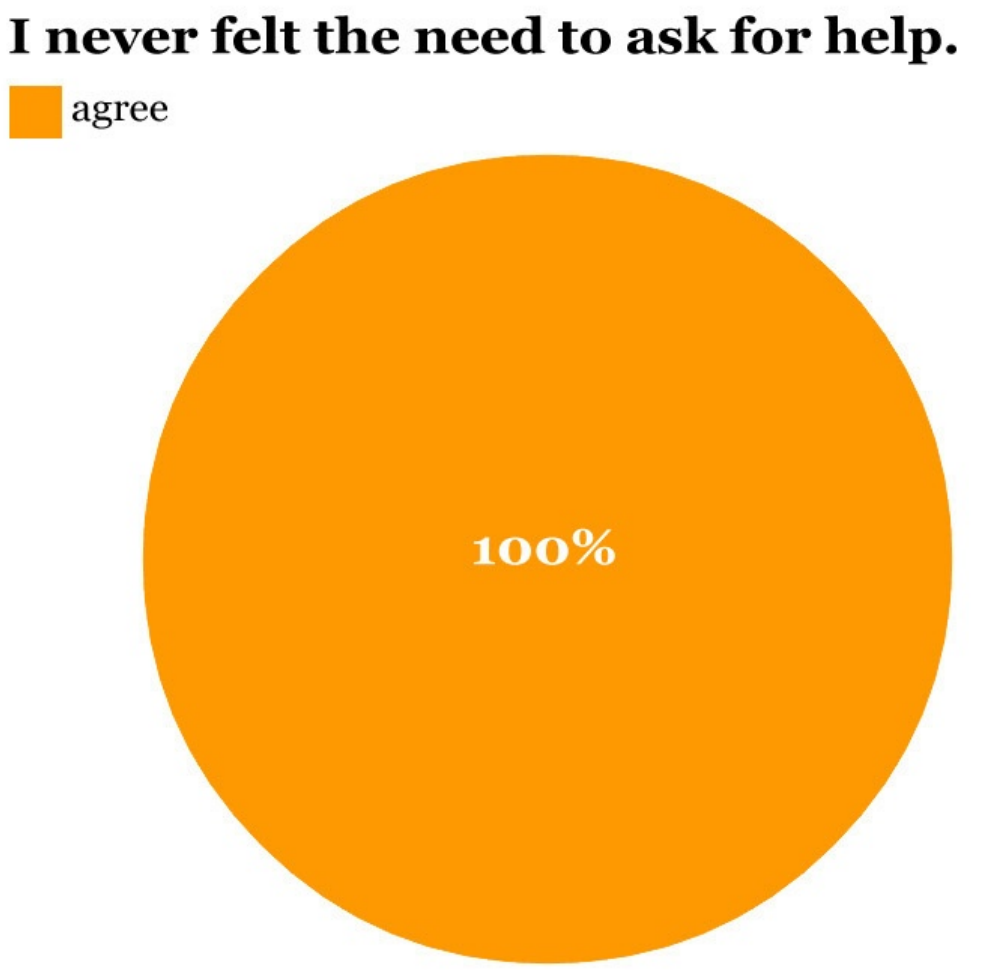